\documentclass[12pt,tightenlines,eqsecnum,floats,aps,amsmath,amssymb,nofootinbib,prd,shownopacs]{revtex4}

\usepackage{amsmath,amssymb,amsfonts,amsthm,amscd}
\usepackage{graphicx}
\usepackage{enumerate} 
\usepackage{colordvi} 
\usepackage{units}
\usepackage{epsfig}

\def\be{\begin{equation}}
\def\ee{\end{equation}}
\def\ba{\begin{eqnarray}}
\def\ea{\end{eqnarray}}

\newcommand{\rcr}{\rho_{\mathrm{crit}}}

\usepackage{enumerate}

\usepackage{colordvi}

\def\R{\mathbb{R}}
\def\S{\mathbb{S}}
\def\G{\mathcal{G}}
\def\I{\mathcal{I}}
\def\L{\rm L}
\def\nu{v}
\def\v{v}
\def\b{$\bullet\,\, $}
\def\f{\frac}
\def\dd{\textrm{d}}

\def\GR{\rm GR}
\def\t{\tilde}

\def\p{p_{\phi}}

\def\Lie{\mathcal{L}}
\def\N{{\cal N}}
\def\lp{{\ell}_{\rm Pl}}
\def\mpl{m_{\rm Pl}}
\def\sp{s_{\rm Pl}}
\def\rp{\rho_{\rm Pl}}
\def\max{\rm max}
\def\avg{\rm avg}
\def\B{\rm B}

\begin{document}
\title{Probability of Inflation in Loop Quantum Cosmology}

\author{Abhay Ashtekar$^1$}\email{ashtekar@gravity.psu.edu}
\author{David Sloan$^{2,1}$}\email{sloan@gravity.psu.edu}
 \affiliation{${}^1$ Institute
for Gravitation and the Cosmos, Physics Department, Penn
State, University Park, PA 16802, U.S.A.\\
${}^2$ Institute for Theoretical Physics, Utrecht University,
The Netherlands. }

\begin{abstract}

Inflationary models of the early universe provide a natural
mechanism for the formation of large scale structure. This success
brings to forefront the question of naturalness: Does a sufficiently
long slow roll inflation occur generically or does it require a
careful fine tuning of initial parameters? In recent years there has
been considerable controversy on this issue \cite{hw1,klm,hw2,gt}.
In particular, for a quadratic potential, Kofman, Linde and Mukhanov
\cite{klm} have argued that the probability of inflation with at
least 65 e-foldings is close to one, while Gibbons and Turok
\cite{gt} have argued that this probability is suppressed by a
factor of $\sim 10^{-85}$. We first clarify that such dramatically
different predictions can arise because the required measure on the
space of solutions is intrinsically ambiguous in general relativity.
We then show that this ambiguity can be naturally resolved in loop
quantum cosmology (LQC) because the big bang is replaced by a big
bounce and the bounce surface can be used to introduce the structure
necessary to specify a satisfactory measure.

The second goal of the paper is to present a detailed analysis of
the inflationary dynamics of LQC using analytical and numerical
methods. By combining this information with the measure on the space
of solutions, we address a sharper question than those investigated
in \cite{klm,gt,as2}: What is the probability of a sufficiently long
slow roll inflation \emph{which is compatible with the seven year
WMAP data}? We show that the probability is very close to $1$.

The material is so organized that cosmologists who may be more
interested in the inflationary dynamics in LQC than in the
subtleties associated with measures can skip that material without
loss of continuity. %
%
%

\bigskip

Key Words: Loop quantum gravity, cosmology, inflation, measures,
probability, WMAP 7 year data.

\end{abstract}


\maketitle

\section{Introduction} \label{s1}

The inflationary paradigm provides a natural mechanism of generating
the seeds of inhomogeneities in the cosmic microwave background
(CMB) which then evolve to the observed, large scale structure of
the universe. The general scenario involves a rather small set of
assumptions: i) Sometime in its early history, the universe
underwent a phase of rapid expansion during which the Hubble
parameter was nearly constant; ii) During this phase, the universe
was well described by a Friedmann, Lema\^itre, Robertson, Walker
(FLRW) solution to Einstein's equations together with small
inhomogeneities which are well approximated by first order
perturbations; iii) Consider the co-moving Fourier mode $k_o$ of
perturbations which has just re-entered the Hubble radius now. A few
e-foldings before the time $t(k_o)$ at which $k_o$ exited the Hubble
radius during inflation, Fourier modes of quantum fields describing
perturbations were in the Bunch-Davis vacuum for co-moving wave
numbers in the range $(k_o,\,\, \sim 200k_o)$; and, iv) Soon after a
mode exited the Hubble radius, its quantum fluctuation can be
regarded as a classical perturbation and evolved via linearized
Einstein's equations. Analysis of these perturbations implies that
there must be tiny inhomogeneities at the last scattering surface
whose detailed features have now been seen in the CMB. Furthermore,
time evolution of these tiny inhomogeneities produces large scale
structures which are in excellent qualitative agreement with
observations. Therefore, even though the assumptions have ad-hoc
elements, they appear to capture a germ of truth, not unlike the
Bohr atom did a hundred years ago.

For definiteness, let us assume a quadratic potential for the
inflaton and supplement our calculations with values of two
parameters provided by the seven year WMAP data \cite{wmap}: the
amplitude $A(t(k_{\star}))$ of the scalar power spectrum
$\Delta_{\rm R}(t(k_{\star}))$ and the scalar spectral index
$n_S(t(k_{\star}))$ at the time the fiducial mode $k_{\star}$ used
by WMAP exits the Hubble radius ($k_{\star} \approx 8.58 k_o$).
These numbers, together with the Friedmann equation, determine the
values of slow roll parameters and \emph{the initial data for the
inflaton and the gravitational field} at time $t(k_{\star})$ to
within posted observational errors. A slow roll follows and dynamics
of perturbations during this epoch directly lead to the spectrum of
inhomogeneities seen in the CMB. We will refer to this inflationary
phase as the \emph{desired} slow roll to distinguish it from other
inflationary phases that may have occurred, e.g., in an even earlier
phase.

The striking success of the scenario brings to forefront an old
issue in a sharper form: Does the desired slow roll inflation occur
generically in a given theoretical paradigm? That is, do generic
dynamical trajectories pass through the neighborhood of the values
of the inflaton and gravitational fields selected by the WMAP data
with its error bars? This would require that the inflaton must have
been significantly high up compared to the minimum of the potential
at the onset of the desired slow roll. How did it get there? Is it
essential to invoke some rare quantum fluctuations to account for
the required initial conditions because the a priori probability for
their occurrence is low? Or, is the desired slow roll inflation
robust in the sense that it is realized in `almost all' dynamical
trajectories of the given theory?

To make these questions precise one needs a well-defined framework
to calculate probabilities of various occurrences \emph{within any
given theory}. A mathematically natural strategy to achieve this
goal was introduced over two decades ago (see e.g.
\cite{ghs,dp,hp}). Recall first that the space $\S$ of solutions to
physically interesting classical systems generally carries the
natural Liouville measure $\dd \mu_{\L}$. The idea was to calculate
\emph{a priori} probabilities using a flat probability distribution
$P(s)=1$ in conjunction with $\dd \mu_{\L}$. More precisely, the a
priori probability of an event $E$ is given by the \emph{fractional}
Liouville volume of $\S$ occupied by the region $R(E)$ consisting of
solutions on which the event $E$ is realized \cite{ghs}. In our
case, then, the a priori probability is given by the fractional
volume occupied by the sub-space of solutions in which the desired
slow roll inflation occurs. Note that this a priori probability
provides only a `bare' estimate and further physical input can and
should be used to provide sharper probability distributions $P(s)$
and a more reliable likelihood. However, a priori probabilities
themselves can be directly useful if they are very low or very high.
In these cases, it would be an especially heavy burden on the
fundamental theory to come up with the physical input that
significantly
alters them. 

However, there is a conceptual obstacle in this calculation: the
total Liouville measure of the space $\S$ of solutions is infinite
\cite{hp}, hence there is an intrinsic ambiguity in the calculation
of relative probabilities \cite{hw2}. But in the observationally
favored k=0 FLRW model, this divergence is a gauge artifact. More
precisely, a gauge group $\G$ acts on $\S$ and, although the
quotient $\S/\G$ ---the space of physically distinct solutions--- is
compact (with boundary), the gauge orbits are non-compact, making
the total volume of $\S$ infinite. It would first appear that the
obvious way to avoid the infinite volume is to work directly with
the space $\S/\G$ of physically distinct solutions. However, as we
will see, the Liouville measure does not naturally project down to
$\S/\G$. Therefore, to calculate probabilities, \emph{one has to
introduce an additional structure.} Because the subtleties
associated with the interplay between the action of the gauge group
and the Liouville measure were not well-understood, the necessity
and importance of this additional step was, apparently, not
appreciated. We will see in section \ref{s3.2} that \emph{there is
an intrinsic ambiguity in carrying out this step within general
relativity.} As a recent analysis of Corichi and Karami \cite{ck}
shows, this ambiguity is directly related to the diverging
conclusions on probability of inflation in general relativity drawn
by Koffmann, Linde and Mukhanov \cite{klm} and Gibbons and Turok
\cite{gt}.

Loop quantum cosmology (LQC) provides a new arena to analyze this
issue because the big bang singularity is naturally resolved and
replaced by a big bounce due to quantum geometry effects
\cite{aps1,aps3,acs,apsv,bp}. We will see in section \ref{s3} that,
thanks to the presence of a canonical bounce time, one can now
naturally resolve the ambiguity in the construction of the measure
on $\S$ with finite total volume. \emph{Were we to try to mimic this
construction in general relativity, we would be led to work at the
singularity in place of the bounce, where the calculation would be
meaningless.} Away from the Planck regime, LQC is virtually
indistinguishable from general relativity. However, in the Planck
regime, there are huge differences and these are crucial in
overcoming the obstacle. With this measure at hand, we can calculate
the a priori probability of the slow roll of the desired type. LQC
dynamics are such that this probability turns out to be greater than
0.999997: Dynamical trajectories starting from almost all initial
data at the bounce surface pass through the phase space region
selected by the WMAP data. Therefore, extreme fine tuning would be
necessary to zero-in on solutions where the desired slow roll does
\emph{not} occur.

Some of the results of our investigation were reported in a Letter
\cite{as2}. Therefore there is an inevitable overlap with \cite{as2}
but there are also key differences. First, in \cite{as2} we analyzed
the likelihood of the occurrence of a slow roll inflation with at
least $\sim 67$ e-foldings in the history of the universe to the
future of the bounce. In this paper we analyze a much \emph{sharper
question:} What is the probability of occurrence of a slow roll with
initial conditions that are compatible with the 7 year WMAP data?
Thus we now focus \emph{only} on that slow roll phase which is
directly relevant to structure formation. Second, numerical
simulations reported in \cite{as2} used values of cosmological
parameters ---the mass of the inflaton and the values of the slow
roll parameters--- from Linde's 2006 review \cite{linde-rev} while
in this paper we use instead the more recent results of the 7 year
WMAP data \cite{wmap}. This accounts for some differences in some of
the detailed numerical results. Finally, and more importantly, our
goal now is broader than that of \cite{as2} in the following sense.
In LQC, the big bounce is followed by a qualitatively new phase of
super-inflation which could have observable consequences. Therefore,
it is important to have a sufficiently detailed account of the new
dynamics from the big bounce to the onset of the desired slow roll.
A second goal of this paper is to provide this analysis. This
detailed description is likely to serve as the point of departure of
further work bridging the Planck era of LQC to the inflationary
paradigm described in the beginning of this section. This bridge
may, for example, provide a better understanding of why the quantum
state of relevant modes is well approximated by the Bunch Davis
vacuum at the onset of inflation, and may even furnish the quantum
gravity corrections to this state \cite{aan}.

The paper is organized as follows. In section \ref{s2} we first
recall the relevant features of LQC and then introduce the phase
space and basic equations. In section \ref{s3} we first introduce
the Liouville measure on the space $\S$ of solutions, discuss the
issue of gauge and obtain the measure on the space $\S/\G$ of
physically distinct solutions in LQC. In section \ref{s4} we discuss
in detail the LQC dynamics from the big bounce to the end of
inflation using a combination of analytical and numerical methods.
Using this information and the measure introduced in section
\ref{s3}, we calculate the a priori probability of obtaining the
desired slow roll. As noted already, in LQC this probability is very
close to 1. Section \ref{s5} summarizes the main results and
compares and contrasts them with related results in the literature.

The material is organized so that cosmologists who may be more
interested in inflationary dynamics of LQC than in the issue of
measures can skip sections \ref{s2.2} and especially \ref{s3}
without loss of continuity.

\section{preliminaries}
\label{s2}

This section is divided into two parts: In the first, we recall the
distinguishing features of LQC that are important to our analysis.
In the second, we present effective LQC equations governing dynamics
of the FLRW model coupled to a scalar field with any potential
(satisfying mild regularity conditions) and introduce the Liouville
measure on the space of solutions.

\subsection{Distinguishing features of LQC}
\label{s2.1}

In LQC, one applies the basic principles of loop quantum gravity
(LQG) \cite{alrev,crbook,ttbook} to simple cosmological models.
Thanks to the quantum geometry underlying LQG, LQC differs from the
older Wheeler-DeWitt theory already at the kinematical level. It
turns out that the Wheeler-DeWitt equation is no longer well-defined
on the new kinematical Hilbert space. Instead, now the quantum
Hamiltonian constraint has to be obtained via a procedure that pays
due attention to the quantum geometry of LQG, in particular, the
area gap $\Delta = 4\sqrt{3}\pi \gamma\lp^2$. (Here $\gamma$ is the
Barbero-Immirzi parameter of LQG, whose value $\gamma \sim 0.24$ is
fixed by black hole entropy calculations \cite{abk,abbdv}.) Somewhat
surprisingly, the resulting dynamics naturally resolves the big bang
and big crunch singularities of general relativity \cite{mb1}.
Exotic matter is not needed; indeed matter fields can satisfy all
the standard energy conditions. Detailed analysis has been carried
out in a variety of models: the k=0, 1\, FLRW space-times with or
without a cosmological constant \cite{aps3,apsv,bp}; Bianchi models
\cite{awe2,awe3,ewe} which admit anisotropies as well as
gravitational waves; and Gowdy models \cite{gowdy} which admit
inhomogeneities, and therefore an infinite number of degrees of
freedom. The FLRW models have been studied most extensively, using
both analytical and numerical methods to solve the exact quantum
equations of LQC \cite{aps3,acs,apsv}. In these models, the big bang
and the big-crunch are replaced by a quantum bounce, which is
followed by a robust phase of super-inflation. Interestingly, full
quantum dynamics, including the bounce, is well-approximated by
certain effective equations \cite{vt,aps3,acs}. These equations
imply that all strong curvature singularities
---including the big rip and sudden-death--- are resolved in FLRW
models with matter satisfying an equation of state of the type
$p=w\rho$ \cite{ps}. (For recent reviews, see \cite{lqcrev}.)

In this paper we will restrict our matter source to be a scalar
field with the standard (positive) kinetic energy and a suitable
potential. Since all the prior discussion of probabilities is based
on general relativity, to facilitate comparison we use effective
equations rather than the full quantum theory. Finally, we will use
the natural Planck units c=$\hbar$=G=1 (rather than $8\pi$G=1, often
employed in cosmology). Planck length will be denoted by $\lp$ and
Planck mass by $\mpl$. The fundamental time unit, $\sp :=
\sqrt{G\hbar/c^5}$, will be referred to as \emph{a Planck second}.

In LQC, it is convenient to encode spatial geometry in a variable
$\v$ proportional to the physical volume of a fixed, fiducial,
cubical cell, rather than the scale factor $a$. The conjugate
momentum is denoted by $b$. These are related to the scale factor
and its conjugate momentum via%
\footnote{In LQG, the basic variable is a triad rather than a
3-metric and in the LQC literature $\v$ is taken to be the oriented
volume which is positive for positively oriented triads and negative
for negatively oriented ones. However, since the change of
orientation is a large gauge transformation, in the classical and
effective theories on can restrict oneself just to positive $\v$. We
have done so for simplicity of discussion.}
\be \label{vb} \v = \f{a^3 V_0}{2 \pi  \gamma} \quad{\rm and} \quad
b=-\f{4 \pi \gamma P_{(a)}}{3 V_0 a^2} \ee
where $V_0$ is the co-moving volume of the fiducial cell, so that
its physical volume is $a^3V_0$. (Thus, the only non-vanishing
Poisson bracket is $\{\v,\, b\} = -2$.) On solutions to
\emph{Einstein's equations}, $b$ is related to the standard Hubble
parameter $H =\dot{a}/a$ via $b= \gamma H$ \cite{acs}. However, LQC
modifies Einstein dynamics and on solutions to the LQC effective
equations we have
\be \label{H} H \,=\, \f{1}{2\gamma\lambda}\, \sin 2\lambda b
\,\,\approx\,\, ({0.93}\,\mpl)\,\, \sin 2\lambda b \ee
where $\lambda^2 := \Delta \approx 5.2 \lp^2$ is the `area-gap' that
sets the discreteness scale of LQC. $b$ ranges over $(0,
\pi/\lambda)$ in LQC and general relativity is recovered in the
limit $\lambda \rightarrow 0$.

Quantum geometry effects modify the geometric, left side of
Einstein's equations. In particular, the Friedmann equation becomes
\be \label{lqc-fe2} \f{\sin^2 \lambda b}{\gamma^2\lambda^2}\,\,
=\,\, \f{8\pi}{3}\, \rho \,\,\equiv\,\, \f{8\pi}{3}\,
\big(\f{{\dot\phi}^2}{2} + V(\phi) \big)\, .\ee
To compare with the standard Friedmann equation $H^2 =
(8\pi/3)\,\rho$, it is often convenient to use (\ref{H}) to write
(\ref{lqc-fe2}) as
\be \label{lqc-fe} \f{1}{9}\,(\f{\dot{\v}}{\v})^2\, \equiv H^2 =
\f{8\pi}{3} \,\,\rho\, \big(1 - \f{\rho}{\rcr}\big) \,  \ee
where $\rcr = 3/8\pi\gamma^2\lambda^2 \approx 0.41 \rp$. By
inspection it is clear from Eqs (\ref{H}) - (\ref{lqc-fe}) that,
away from the Planck regime ---i.e., when $\lambda b \ll 1$, or,
$\rho \ll \rcr$--- we recover classical general relativity. However,
modifications in the Planck regime are drastic. The main features of
this new physics can be summarized as follows. 
\medskip

\b In general relativity, the Friedmann equation implies that if the
matter density is positive, $\dot{a}$ cannot vanish. Therefore every
solution represents \emph{either} a contracting universe \emph{or}
an expanding one. By contrast, the LQC modified Friedmann equation
(\ref{lqc-fe}) implies that $\dot{\v}$ vanishes at $\rho=\rcr$. This
is the quantum bounce. To its past, the solution represents a
contracting universe with $\dot{\v} <0$ and to its future, an
expanding one with $\dot{\v} >0$.

\b As is customary in the literature on probabilities, let us ignore
the exceptional de Sitter solutions. On all other solutions $b$ is
monotonically non-increasing, evolving from from $b = \pi/\lambda$
in the infinite past to $0$ in the infinite future. Eqs
(\ref{lqc-fe2}) and (\ref{lqc-fe}) imply that $b=\pi/2\lambda$ at
the bounce. Thus, each solution undergoes precisely one bounce.

\b In contrast to general relativity, the Hubble parameter $H
=\dot{\v}/3\v$ is no longer monotonic in LQC. It \emph{vanishes} at
the bounce while in general relativity it diverges at the
singularity and is large in the entire Planck regime. In LQC, $H$ is
bounded above, $|H| \lesssim 0.93\,\mpl$, and achieves its upper
bound in every solution at the end of super-inflation.

\b If the potential $V(\phi)$ is bounded below, say $V \ge
V_o$, then it follows from (\ref{lqc-fe2}) that ${\dot\phi}^2$
is bounded by $2\rcr - 2V_o$. If $V$ grows unboundedly for
large $|\phi|$, then $|\phi|$ is also bounded. For example, for
$V= m^2\phi^2/2$, we have $m|\phi|_{\rm max} = 0.90 \,\mpl^2$.

\b When the potential is bounded below, $|\dot{H}|$ is bounded above
by $10.29 \,\mpl^2$. The Ricci scalar ---the only non-trivial
curvature scalar in these models--- is bounded above by $31\mpl^2$.
Thus, physical quantities which diverge at the big bang of general
relativity cannot exceed certain finite, maximum values in LQC. One
can also show that if $v \not=0$ initially, it cannot vanish in
finite proper time along any solution. Thus, \emph{the LQC solutions
are everywhere regular irrespective of whether one focuses on matter
density, curvature or the scale factor.} \\

\subsection{The Liouville measure}
\label{s2.2}

The full set of space-time equations of motion can be written in
terms of $\v(t),\phi(t)$. These variables are subject to the
constraint (\ref{lqc-fe}) and evolve via:
\ba \label{dyn} &&\ddot{\v} = \f{24\pi \v}{\rcr}\,\big[(\rho
-V(\phi))^2 + V(\phi) (\rcr-V(\phi))\big]\\
&&\ddot\phi + \f{\dot{\v}}{\v}\, \dot\phi + V_{,\phi} =0\, . \ea
To calculate probabilities, in section \ref{s3} we have to equip the
space $\S$ of solutions to these equations with a natural measure.
As a first step in that procedure, we will now obtain a phase space
formulation of these equations. The phase space and equations of
general relativity can be recovered by taking the limit $\lambda \to
0$ (which in particular implies $\rcr \to \infty$).

The phase space $\Gamma$ consists of quadruplets $(\v,b;\,
\phi, \p)$, where $\p$ is given by; $\p = 2\pi\gamma\lp^2\,\v\,
\dot\phi$.\, The variables $\phi,\p$ range over the entire real
line, $\v$ over the positive half of the real line, while $b
\in [0, \pi/2\lambda]$ (since we focus only on the post-bounce
branches of solutions). Thus, the symplectic 2-form is given by
\be \label{Omega} \Omega = \dd\phi \wedge \dd\p + \f{1}{2} \dd b
\wedge \dd\v \ee
Hence the Liouville measure on $\Gamma$ is simply $2\dd\mu_{\L} =
\dd\phi\, {\dd}\p\,{\dd} {b}\, {\dd} {\v}$.

The LQC Friedmann equation implies that these variables must lie on
a constraint surface $\bar\Gamma$ defined by
\be \label{hc}  C \equiv -\f{3\v}{4\gamma\lambda^2}\, \sin^2\lambda
b \, + \,\f{\p^2}{4\pi\gamma\v} + 2\pi\gamma\v\, V(\phi)\,\approx\,
0\, . \ee
They evolve via
\ba\label{evo}
 \dot{\v} &=& \f{3v}{2\gamma}\, \f{\sin 2\lambda b}{\lambda},
 \quad\quad
 \dot{b} = - \f{\p^2}{\pi \gamma \v^2},\\
 \dot\phi &=& \f{\p}{2\pi\gamma \v},  \quad\quad\quad\quad
 {\dot p}_{(\phi)} = -2\pi\gamma \v\, V_{,\phi}\, . \ea
As is well-known, the space of solutions $\S$ is naturally
isomorphic to a gauge-fixed surface, i.e., a 2-dimensional surface
$\hat\Gamma$ of $\bar\Gamma$ which is intersected by each dynamical
trajectory once and only once. Since $b$ is monotonic in each
solution, an obvious strategy is to choose for $\hat\Gamma$ a
2-dimensional surface $b= b_o$ (a fixed constant) within
$\bar\Gamma$.

It is straightforward to pull-back the symplectic structure to this
2-dimensional gauge-fixed surface $\hat\Gamma$. Using the constraint
(\ref{hc}), it is convenient to coordinatize $\hat\Gamma$ using $\v,
\phi$ and express the pulled-back symplectic structure in terms of
them:
\be \label{Omegahat} \hat\Omega = \big[\f{3\pi}{\lambda^2}
\sin^2\lambda b_o - 8\pi^2 \gamma^2 V(\phi)\big]^{\f{1}{2}}
\,\,\dd\phi \wedge \dd\v \ee
This 2-form provides provides a Liouville measure $\dd\hat\mu_{\L}$
on $\hat\Gamma$ and hence of the space $\S$ of solutions to the
effective equations, given simply by
\be \label{muhat} \dd\hat\mu_{\L} = \big[\f{3\pi}{\lambda^2}
\sin^2\lambda b_o - 8\pi^2 \gamma^2 V(\phi)\big]^{\f{1}{2}}
\,\,\dd\phi\, \dd\v\, . \ee
The most natural choice in LQC is to set $b_o = \pi/2\lambda$
so that $\bar\Gamma$ is just the `bounce surface'. We will make
this choice because it also turns out to be convenient for
calculations. However, since the dynamical flow preserves the
symplectic structure on $\Gamma$, this measure on $\S$ is
insensitive to the choice of $b_o$ used in gauge fixing.


\section{Probability considerations}
\label{s3}


Recall from section \ref{s1} that the a priori probability of
occurrence of any event $E$ is to be given by the fractional
volume of the region $R(E)$ in $\S$ spanned by solutions in
which E occurs:
\be \label{Prob} P(E)=\f{\int_{R(E)} \dd\hat{\mu}_{\rm L}}{\int_\S
\dd\hat\mu_{\rm L}}\, . \ee
Since the Liouville measure is purely kinematical, this method of
calculating a priori probabilities realizes Laplace's
\emph{principle of indifference} \cite{psdl}. This interpretation
can be made explicit as follows. Physical input can provide a
(non-negative) probability density function $\rho(s)$ on $\S$
satisfying the normalization condition [$\int_{\S} \rho\,
\dd\hat{\mu}_{\rm L}]/ [\int_{\S}\dd\hat{\mu}_{\rm L}] =1$. The
corresponding (more refined) probability of occurrence of $E$ is
then given by
\be P_{\rho}(E) = \f{\int_{R(E)}\, \rho(s)\,\,
\dd\hat{\mu}_{\rm L}}{\int_\S \rho(s)\,\, \dd\hat\mu_{\rm L}}\,
. \ee
Now, one can quantify the information contained in $\rho(s)$
via
\be I_\rho = \f{\int_\S \rho(s)\, \ln \rho(s)\, \dd\hat\mu_{\rm
L}}{\int_\S \rho(s)\,\, \dd\hat\mu_{\rm L}}\, . \ee
Since $I$ is minimized by $\rho(s) =1$, (\ref{Prob}) is the
\emph{a priori} probability, free of any input or bias. As
mentioned in section \ref{s1}, these `bare' probabilities are
useful when they are extremely low or extremely high. In these
cases it is a heavy burden on any theory to provide sufficient
information to significantly overcome this bare probability.

In section \ref{s3.1} we illustrate these ideas using a
2-dimensional simple harmonic oscillator constrained to have a fixed
energy. In this case the solution space $\S$ is compact. Since the
total Liouville volume of $\S$ is finite, one can directly implement
the ideas outlined above to calculate various probabilities. In
section \ref{s3.2} we turn to LQC. In this case, $\S$ is
non-compact. But we will see that this non-compactness can be
directly attributed to the action of a gauge group $\G$ on $\S$.
Therefore, it is now natural to work with the space $\t{S} := \S/\G$
of physically distinct solutions which is compact. However, there is
a subtlety which makes the calculation of desired probabilities
ambiguous in general relativity. We discuss this problem and show
that it has a natural resolution in LQC.

Cosmologists who are more interested in the LQC dynamics than in the
issue of measure on $\S$ can skip section \ref{s3} without loss of
continuity.

\subsection{Harmonic oscillator}
\label{s3.1}

Since we encounter a constrained Hamiltonian system both in
general relativity and (effective) LQC, let us begin with a
simple example with this feature to illustrate the procedure of
calculating probabilities. Consider a 2-dimensional simple
harmonic oscillator constrained to have a fixed energy $E$. Let
us fix the mass and spring constant to unity for simplicity.
The topology of our phase-space $\Gamma$ is then
$\mathbb{R}^4$, in which we can choose canonical coordinates
$x_1,p_1;\,x_2, p_2$, with symplectic structure:
\be \omega = \dd x_1 \wedge \dd p_1 + \dd x_2 \wedge \dd p_2
\ee
The Hamiltonian constraint is:
\be H :=\f{1}{2} (x_1^2+p_1^2+x_2^2+p_2^2) =E \ee
The constrained surface $\bar\Gamma$ is the 3-sphere with radius
$\sqrt{2E}$ in $\mathbb{R}^4$ and thus compact. The Hamiltonian
vector field is given by
\be X_H = x_1 \f{\partial}{\partial p_1} - p_1
\f{\partial}{\partial x_1} + x_2\f{\partial}{\partial p_2} -
p_2 \f{\partial}{\partial x_2} \ee
The vector field is of course tangential to $\bar\Gamma$ and
its orbits provide a Hopf fibration: each orbit is an $S^1$
fiber with the base space being $S^2$. This base space
represents the space $\S$ of solutions of the constrained
system under consideration.

It is convenient to use a set of coordinates adapted to this
Hopf fibration:
\be x_1+ip_1 = \sqrt{2E}\, e^{i\xi_1} \sin \eta \equiv z_1
\quad {\rm and} \quad x_2+ip_2 = \sqrt{2E}\, e^{i\xi_2} \cos
\eta \equiv z_2 \ee
where $\eta \in (0,\pi/2)$ and $\xi_1, \xi_2 \in (0, 2\pi)$.
Then $\bar\Gamma$ is given by $|z_1|^2 + |z_2|^2 = 2E$ and the
angles $\eta, \xi_1, \xi_2$ provide a a natural set of
intrinsic coordinates on it. The pull-back $\bar\Omega$ of
$\Omega$ to $\bar\Gamma$ can now be expressed as
\be \bar\Omega = 2E\, \sin 2\eta\,\,\, \dd \eta\wedge \dd \xi^-
\ee
and the restriction of the Hamiltonian vector field to
$\bar\Gamma$ is given by
\be \bar{X}_H = 2 \f{\partial}{\partial \xi^+} \ee
where we have set $2\xi^\pm = \xi_1 \pm \xi_2$. As expected,
$\bar\Omega$ is degenerate and the degenerate direction is
precisely $\bar{X}_H$. It is obvious that $\eta, \xi^-$ and
$\bar\Omega$ are Lie dragged by $\bar{X}_H$. Hence they have
unambiguous projections to the 2-sphere of orbits of
$\bar{X}_H$, i.e., to the space $\S$ of solutions. The total
Liouville volume of $\S$ is given by:
\be \int_\S \bar\Omega\, =\, 2E \int_0^{\pi/2}\! \dd \eta\,
\sin 2\eta \,\int_{0}^{2\pi}\!  \dd\xi^- \, = \, 2\pi (2E)\, .
\ee

We can now use the Liouville measure $\dd\mu_{\rm L} = (2E)
\sin 2\eta\, \dd\eta\dd\xi^- $ on $\S$ to calculate the a
priori probabilities of physical events of interest. Consider
example, the total angular momentum
\be J := x_1 p_2 - x_2 p_1 = -E\, \sin 2\eta\, \sin 2\xi^-\ee
which is a Dirac observable and thus projects down from
$\Gamma$ to $\S$ unambiguously. Another Dirac observable is the
eccentricity of the orbit in $x_1,x_2$ plane defined in each
solution, $e= {r_{min}}/{r_{max}}$ where $r^2 = x_1^2 + x_2^2$.
One can show that $e$ is determined completely by the ratio
$|J|/E$:
\be  e=\f{r_{min}}{r_{max}} =
\sqrt{\f{1-\sqrt{1-\f{J^2}{E^2}}}{1+\sqrt{1-\f{J^2}{E^2}}}}\, .
\ee
To gain insight into the likelihood of various values for both
observables, it is natural to ask, e.g., for the a priori
probability that $|J|/E$ is larger than a given number $f \in
[0,1]$. We obtain :
\be P\big(\, (|J|/E) >f\, \big) =\f{1}{4\pi E} \int_A \bar\Omega =
1-f \ee
where $A$ is the region such that $\sin 2\eta \sin 2\xi^- > f$. We
find that the probability for $e>1/2$ is $P_{|J|/E > 16/25} =
1-16/25 = 9/25$ and the probability for $e>3/4$ is only $49/625
\approx 0.08$. We can therefore conclude that almost circular orbits
($e \approx 1$) are rare according to our measure, and thus if the
orbit of a particle in this system were observed to be very close to
circular, further physical explanation would be required.

Finally, in this analysis we identified the solution space $\S$
with the quotient of $\bar\Gamma$ by the orbits of the
Hamiltonian vector field. In general relativity and LQC it is
more convenient to gauge fix. In the present case, because the
orbits of $X_H$ provide a Hopf fibration of $\bar\Gamma$ a
global gauge fixing is not available. Still, for calculating
probabilities, we can ignore sets of measure zero and
parameterize $\S$ by points of a 2-dimensional surface
$\hat\Gamma$ given by $\xi^+ = {\rm const}$. For physical
questions such as the ones discussed above, the resulting a
priori probabilities are insensitive to the choice of the
constant and the results are the same as those obtained above.

\subsection{Measures for general relativity and LQC}
\label{s3.2}

Recall from section \ref{s2} that the phase space $\Gamma$ of LQC is
naturally coordinatized by $(\v, b; \phi,\p)$ where $v$ takes values
in the positive half line, $\phi,\p$ on $\R$ and $b\in (0,\,
\pi/\lambda)$. These phase space variables are subject to the
Hamiltonian constraint (\ref{hc}) of effective LQC and dynamics of
this theory is generated by the Hamiltonian vector field $X_C$ of
the constraint function $C$:
\be \label{hamvf} X_C = \big(\f{3v}{2\gamma}\, \f{\sin 2\lambda
b}{\lambda}\big)\f{\partial}{\partial v} - \big(\f{\p^2}{\pi \gamma
\v^2}) \f{\partial}{\partial b} + \big(\f{\p}{2\pi\gamma \v}\big)
\f{\partial}{\partial \phi} - \big(2\pi\gamma \v\, V_{,\phi}\big)
\f{\partial}{\partial \p}\, . \ee
There is a gauge symmetry which will play an important role in our
probability considerations. Since $v$ is, by definition, the
physical volume of a fiducial cell and the choice of this cell is
arbitrary, physics does not change under the transformation
\be \label{gauge} (v,b; \phi,\p) \quad \longrightarrow \quad (\alpha
v, b; \phi, \alpha\p) \ee
which corresponds merely to a rescaling of the fiducial cell by a
positive number $\alpha$ (recall: $\p = 2\pi\gamma\lp^2\,
v\,\dot\phi$). Thus the phase space $\Gamma$ is acted upon by the
action of a gauge group $\G$ generated by (\ref{gauge}). (This gauge
freedom is equivalent to that in rescaling the scale factor in the
more familiar Wheeler-DeWitt phase space.) At an infinitesimal level
this rescaling defines a vector field
\be G = v\, \f{\partial}{\partial v} + \p\, \f{\partial}{\partial
\p}\, . \ee
Under this rescaling the constraint function (\ref{hc}) and the
symplectic structure (\ref{Omega}) are merely rescaled, $\Lie_{G} C
= C$ and  $\Lie_G\Omega = \Omega$, such that the Hamiltonian vector
field (\ref{hamvf}) is left invariant, $\Lie_{G} X_C =0$. Hence, as
one would expect of gauge transformations, $\G$ descends to the
space $\S$ of solutions. These considerations hold both for
effective LQC and general relativity.

Next, let us find a suitable parametrization of $\S$. Note first
that we can solve (\ref{hc}) for $\p$:
\be \label{pphi}\p = \pm \, \big[\f{3\pi \v^2}{\lambda^2}\,
\sin^2\lambda b - 8\pi^2\gamma^2 \v^2 V(\phi)\big]^{1/2}. \ee
Without loss of generality we can restrict ourselves to the positive
branch $\p \ge 0$ because identical considerations will apply to the
negative branch. Thus the constraint surface $\bar\Gamma$ is
coordinatized by $(\v,b,\phi)$. Since the dynamical trajectories lie
in $\bar\Gamma$, the space of solutions $\S$ is 2-dimensional and
naturally isomorphic to any gauge-fixed surface $\hat\Gamma$ in
$\bar\Gamma$ which is intersected by each dynamical trajectory once
and only once. Since $b$ is monotonic along dynamical trajectories,
we can take $\hat\Gamma$ to be the 2-dimensional surface $(v, b=b_o,
\phi)$ for any fixed constant $b_o$ in the domain $(0, \pi/\lambda)$
of $b$.%
\footnote{It is then convenient to choose the freedom in rescaling
the constraint ---i.e., in the choice of the lapse function--- to
make $b$ the affine parameter of the Hamiltonian vector field, so
that $X_C$ now maps each $b={\rm const}$ surface to another $b={\rm
const}$ surface.}
The Liouville measure $\dd\hat\mu_{\L}$ on $\hat\Gamma$, and hence
on $\S$, is then given by (\ref{muhat}):
\be \dd\hat\mu_{\L} = \big[\f{3\pi}{\lambda^2} \sin^2\lambda b_o -
8\pi^2 \gamma^2 V(\phi)\big]^{\f{1}{2}} \,\,\dd\phi\, \dd\v\, . \ee
Since it is induced by the symplectic structure, the
$\dd\hat\mu_{\rm L}$ volume of any region on the $b=b_o$ section is
the same as that of its image on any other $b={\rm const}$ surface
under the Hamiltonian flow $X_C$. In this sense, $\dd\hat\mu_{\rm
L}$ is insensitive to the choice of $b_o$. The analogous structures
$\Gamma_{\GR}, \bar\Gamma_{\GR}, \hat\Gamma_{GR}$ and $\dd
\hat\mu_{\L}^{\GR}$ in general relativity are obtained by taking the
limit $\lambda \to 0$. In particular, in this limit $b \to H$, the
Hubble parameter. Hence in general relativity one can naturally fix
the gauge by setting $H=H_o$ for any positive constant $H_o$.

The essential problem in both cases is that the spaces of solutions
$\S$ and $\S_{\GR}$ are non-compact and their volume with respect to
$\dd \hat\mu_{\L}$ and $\dd \hat\mu_{\L}^{\GR}$ is infinite.
Therefore we cannot directly use the procedure of section \ref{s3.1}
to calculate probabilities of events of interest. Let us restrict
ourselves to positive potentials $V(\phi)$ ---such as the quadratic
and quartic ones which are often analyzed in detail--- which diverge
at $\phi=\pm\infty$. Then the constraint (\ref{hc}) (and its GR
analog, the Friedmann equation) immediately implies that \emph{the
range of $\phi$ is bounded} on $\hat{\Gamma}$. Thus, the
non-compactness of $\hat\Gamma$ and $\hat\Gamma_{\GR}$ arises only
because $v$ ranges over the entire positive real line. However, as
we saw in the beginning of this sub-section, the transformation
$(v,\phi) \to (\alpha v,\phi)$ is a gauge motion. Thus, the total
Liouville measure of $\S$ is infinite simply because each orbit of
the gauge group $\G$ is infinitely long.

Let us spell out the interaction between the gauge group and the
space of solutions $\S$ explicitly. Since $\Lie_{G} C =0$, the gauge
vector field $G$ is tangential to $\bar\Gamma$. Furthermore since
$\Lie_{G} b =0$ (and $\Lie_{G} H=0$ in general relativity), $G$ is
also tangential to the gauge-fixed surface $b=b_o$ in effective LQC
(and to the gauge-fixed surface $H=H_o$ in general relativity).
Thus, $G$ is tangential to $\hat\Gamma$ (and $\hat\Gamma_{\GR}$),
i.e., the action of the gauge group $\G$ naturally respects our
parametrization of the space $\S$ of solutions using $\hat\Gamma$.
The space of \emph{physically distinct} solutions is the quotient
$\t\S:= \S/\G$, coordinatized just by $\phi$ which takes values on a
compact interval, say $(\phi_{\rm min}, \phi_{\rm max})$. Since we
are interested in gauge invariant questions, regions $R(E)$ of $\S$
consisting of solutions in which \emph{physical} events ---such as
the desired slow roll inflation--- occur contain whole orbits of
$\G$ and project down unambiguously to $\t\S$. To calculate the
probability of such events, then, it would suffice to have available
a natural measure on $\t\S$. Since $\t\S$ is a closed interval, one
would expect a natural construction to provide it with a finite
total measure, making the calculation of desired probabilities
well-defined as in section \ref{s3.1}.

Thus, the natural strategy is to attempt to introduce a volume
element on $\t\S = \hat\Gamma/\G$ starting from the symplectic
structure $\hat\Omega$ on $\hat\Gamma$.%
\footnote{A straightforward push-forward of the Liouville measure
$\hat\mu_{\L}$ on $\S = \hat\Gamma$ is not useful because it would
assign to any interval on $\hat\Gamma/\G$ the same volume as its
inverse image in $\hat\Gamma$ which is infinite.}
The obvious candidate for such a volume element is the natural
1-form $\hat\omega_\alpha =\hat\Omega_{\alpha\beta} G^\beta$ which
is transverse to the gauge direction $G^\alpha$. Unfortunately,
since $\Lie_{G} \hat\omega_\alpha= (\Lie_{G}\,
\hat\Omega_{\alpha\beta}) G^\beta = \hat\omega_\alpha \not= 0$, this
1-form $\hat\omega_\alpha$ does \emph{not} project down to $\t\S$.
Therefore to implement this strategy one has to introduce an
additional structure. Let us fix a line $\S_o$ in $\S = \hat\Gamma$
given by $v=v_o$ for some constant $v_o$, which is naturally
isomorphic to $\t\S = \hat\Gamma/\G$.%
\footnote{The use of a more general line, $v=f(\phi)$, would
correspond to giving the gauge orbits a $\phi$ dependent regularized
volume. This would introduce an ad-hoc new input in the
construction. With $v=v_o$, the regularized volume of each orbit is
the same and, as explained below, it drops out in the calculation of
probabilities of physical events.}
One can then just pull-back the 1-form $\hat\omega$ to $\S_o$ and
perform integrations there. Expression (\ref{Omegahat}) of
$\hat\Omega$ implies that the total volume of $\S_o$ is given by
\be \label{vol}\int_{\S_o} \hat\omega = \int_{-\phi_{\rm
max}}^{\phi_{\rm max }}\, \big(\f{3\pi}{\lambda^2} \sin^2 \lambda
b_o - 8\pi^2\gamma^2 V(\phi)\big)^{\f{1}{2}}\, v_o\,\dd\phi\, , \ee
which is manifestly finite. Finally, a change in the value of $v_o$
would  simply rescale $\hat\omega$ by a \emph{constant}. Since the
relative probability $P(E)$ of the occurrence of a physical event
$E$ is given by
\be \label{prob} P(E) = \f{\int_{\I(E)}\, \hat\omega}{\int_{\S_o}
\hat\omega}\, , \ee
where $\I(E)$ is the intersection of the $v=v_o$ line and the region
$R(E)$ on which the property $E$ is realized, clearly, $P(E)$ is
independent of the choice of $v_o$ made above.

Thus, it would appear that we have satisfactorily eliminated the
spurious infinity in the total Liouville measure and arrived at a
natural procedure to calculate desired probabilities. However, there
is an important subtlety which makes the procedure ambiguous. Recall
that in order to arrive at $\hat\omega$ we made \emph{two} gauge
choices: We set $b=b_o$ (or, $H=H_o$ in GR) and then $v=v_o$. The
symplectic structure and the 1-form $\hat\omega$ on $\S$ is
preserved by the Hamiltonian flow and (with the appropriately chosen
lapse) the flow maps each of our 2-dimensional gauge fixed surfaces
$b={\rm const}$ to another such surface. However, under this flow,
the data $(v=v_o, b=b_o;\,\, \phi,\p=\p(\v_o,b_o,\phi))$ with
constant $v$ on the $b=b_o$ slice is \emph{not} mapped to data with
constant $v$ at another slice, say $b=b_1$. (Here
$\p(\v_o,b_o,\phi)$ is the solution (\ref{pphi}) of the constraint).
Therefore the final step, where we carry out the integration on the
$v=v_o$ line, of our procedure sensitively depends on our initial
choice of $b_o$ (or $H=H_o$ in general relativity). The question
then is: Is there a canonical choice $b_o$ one can make in LQC (or,
$H_o$ in general relativity)?

In general relativity, the answer is in the negative: there is no
preferred value of the Hubble parameter or a canonical instant of
time in any solution. Therefore the calculation of probabilities has
an intrinsic ambiguity. Although there are also other differences
(discussed in section \ref{s5}), this ambiguity lies at the source
of dramatically different predictions on the probability of
inflation because, in effect, by choosing very different values of
$H_o$, one can arrive at very different measures for computing
probabilities. In terms of these considerations, Gibbons and Turok
chose a low value of $H_o$ and found that the sub-space of solutions
admitting a sufficiently long slow roll occupies an extremely tiny
relative volume of $\S$ with respect to the resulting measure on
$\S$. A higher value of $H_o$ ---as was advocated in the early
literature (see, e.g. \cite{bkgz})--- increases this probability
very substantially. In LQC, on the other hand, the bounce surface
provides a canonical `time' instant, i.e., a natural value for
$b_o$, namely $b_o=\pi/2\lambda$ where the matter density attains
its maximum value. The corresponding surface in general relativity
would be the big bang singularity and, unfortunately, we cannot use
the `data at the big bang' to parameterize the space $\S$ of
solutions.
\footnote{A common suggestion that one should carry out the
calculation when the Hubble parameter is of Planck scale, rather
than infinite, comes close to the natural strategy in LQC but the
suggestion is only qualitative and, strictly speaking, quite
different because in LQC the Hubble parameter vanishes at the bounce
surface.}

To summarize, as discussed in the early literature \cite{ghs,dp,hp},
the Liouville measure on the phase space naturally descends to the
space $\S$ of solutions but the total $\dd\hat\mu_{\rm L}$ measure
of $\S$ is infinite. In the observationally favored k=0 models
this infinity is physically spurious because it arises only because
the length of gauge orbits is infinite. For physical questions one
can work with the quotient $\S/\G$ which is just a bounded interval
of the real line both in general relativity and LQC. However,
although the Hamiltonian vector field $X_C$ is invariant under the
action of the gauge group $\G$, the symplectic structure is not.
Consequently, the Liouville measure fails to naturally descend to a
measure (with finite total volume) on $\S/\G$. To introduce such a
measure, an additional structure is needed. Because of the presence
of the bounce, this structure is naturally available in LQC. In
general relativity, by contrast, there does not appear to exist a
natural way to introduce the required structure; the analog of the
bounce surface would be the big bang singularity itself!
Consequently, in general relativity, the calculation of a priori
probabilities of various physical events along the lines of
\cite{ghs,dp,hp} is intrinsically ambiguous.

\emph{Remark:} In LQC, the final result can be stated directly in
terms of the Liouville measure $\dd\hat\mu_{\L}$ on $\S=\hat\Gamma$.
The regions of interest $R(E)$ spanned by solutions in which a
physical event $E$ occurs contain whole orbits of the gauge group
$\G$:\, $R(E) = I\times \R^+$ where $I$ is a closed interval in
$[-\phi_{\rm max},\, \phi_{\rm max}]$ and $\R^+$ denotes the $v$
axis ($\phi$ and $\v$ evaluated at the bounce surface
$b=\pi/2\lambda)$. The associated probability $P(E)$ is given by
first introducing a `slab' $I_{v_o}$ in the 2-dimensional space
$\hat\Gamma$, bounded by $\v= \v_o$ and $\v= 1/\v_o$ with
$0<\v_o<1$, and then taking the limit $\v_o\to 0$:
\be  P(E)\,\,= \,\, \lim_{v_0 \rightarrow 0}\, \f{\hbox{\rm
Liouville Volume of}\,\, [I \times I_{v_0}]}{\hbox{\rm Liouville
Volume of}\,\, [I_{\rm total}\times I_{v_0}]}\,\,\,  \ee
This is the expression that was used in \cite{as2}.

\section{Inflationary dynamics in LQC}
\label{s4}

The event $E$ of interest to us is the presumed slow roll that is
compatible with the WMAP data. With a canonical measure on the space
$\S/\G$ of physically distinct solutions of LQC at hand, we can now
ask for the relative volume occupied by the region $R(E)$ consisting
of solutions in which $E$ occurs. In this section we will first
characterize this subspace by analyzing in detail the post-bounce
dynamics of LQC and then calculate its relative volume.

In section \ref{s4.1} we collect useful facts about the desired slow
roll that are used throughput the subsequent subsections. Solutions
of effective LQC equations can be naturally identified with their
initial data at the bounce surface. It turns out that the
qualitative features of the relevant dynamical evolution are largely
dictated by the ratio of the kinetic to the potential energy in this
initial data. In particular, as we will see in section \ref{s4.3},
only those dynamical trajectories for which the kinetic energy
overwhelmingly dominates fail to lie in $R(E)$. To bring out this
point, in section \ref{s4.2} we will study in detail the solutions
in which the kinetic energy completely overwhelms the potential
energy at the bounce. In section \ref{s4.3} we first briefly discuss
dynamics in other solutions and then show that except in the case of
\emph{very extreme} kinetic energy domination at the bounce, the
solutions necessarily lie in $R(E)$. In section \ref{s4.4} we find a
bound on the relative volume of $R(E)$ which shows that the
probability of seeing the desired phase of slow roll inflation is
extremely close to $1$ in LQC. While the detailed discussion uses a
quadratic potential, we argue that the qualitative result is likely
to hold much more generally.

\subsection{The desired slow roll}
\label{s4.1}

Let us begin by recalling the slow roll parameters that are used in
the discussion of the WMAP data. There are two conceptually distinct
sets of parameters \cite{lpb}
\be \label{parameters}
\epsilon = -\f{\dot{H}}{H^2},\quad \eta =
\f{\ddot{H}}{\dot{H}{H}}\qquad {\rm and} \qquad \epsilon_V :=
\f{1}{16\pi}\,\big(\f{V^\prime}{V}\big)^2\, \,\mpl^2,\quad
 \eta_V =  \f{V''}{8\pi V}\, \,\mpl^2\, . \ee
Smallness of the first set ensures that the Hubble parameter is
changing very slowly in the dynamical phase under consideration.
(Sometimes one uses the symbol $\epsilon_H$ for $\epsilon$ and
$\eta_H$ for $\eta$ to highlight this.) Inflation ---i.e.,
accelerated expansion--- is characterized by $\epsilon <1$. For slow
roll, on the other hand, we need $\epsilon \ll 1$ and $\eta \ll 1$.
As is common in the literature, in our study of slow roll, we will
keep terms of order 1 in these parameters but ignore higher order
terms. For concreteness, we will end the slow roll phase when
$\epsilon=0.1$ so that the errors will be less than a percent. In
theoretical investigations one often uses the second set of slow
roll parameters which is tailored to the properties of the
potential. During the epoch in which general relativity is an
excellent approximation, the two sets are closely related by
dynamics. In particular,
$\epsilon_V=[(1+\eta/3)/(1-\epsilon/3)]^2\,\, \epsilon$. Hence the
difference between $\epsilon$ and $\epsilon_V$ is of second order in
$\epsilon,\, \eta$. Finally, these parameters are best suited for
studying dynamics during the general relativity era. In full LQC, on
the other hand, $\dot{H}=0$ at the end of super-inflation but the
super-inflation phase is far from being quiescent if the bounce is
kinetic energy dominated. A better definition of the first slow roll
parameter is $\epsilon= (3 {\rm KE})/(2\rho)$. In general
relativity, this definition agrees with (\ref{parameters}) but with
this definition the slow roll condition $\epsilon \ll 1$ is violated
even at the end of super-inflation if the bounce is kinetic energy
dominated. In this paper we will use the standard definition
(\ref{parameters}) because all our slow roll considerations will
refer only to the general relativity epoch.

Next, let us recall the constraints imposed by the WMAP data on
scalar perturbations. The data is tailored to the co-moving wave
number $k_{\star}$ given by \cite{wmap}
\be \f{k_{\star}}{a_o}  =  2\times 10^{-3}\, {\rm Mpc}^{-1}\qquad
{\rm or}\qquad k_{\star} = 8.58\, k_o  \ee
where $a_o$ refers to the scale factor today and, as before, $k_o$
refers to the wave number that has just re-entered the Hubble radius
today. (It is only the combination $k_{\star}/a_o$ that has direct
physical meaning; $2\pi a_o/k_o$ is the physical wave length of this
reference mode today.) Within inflationary models, the data
constrains initial values of fields describing the homogeneous
isotropic background at time $t(k_{\star})$ i.e., \emph{the time at
which the mode $k_{\star}$ exits the Hubble radius during
inflation.} In the rest of this subsection, we will make these
constraints explicit.

The amplitude $A(t(k_{\star}))$ of the scalar power spectrum
$\Delta_{\rm R}^2 (k_{\star})$ at this wave number is given by:
\be  A(t(k_{\star})) = \f{H^2(t(k_{\star}))}{\pi \epsilon\,
(t(k_{\star}))\,\mpl^2} = 2.43\times 10^{-9} \ee
and the scalar spectral index $n_S(k_{\star})$ is given by
\be n_s(t(k_{\star})) = 1- \f{\dd \ln \Delta_{\rm R}^2}
 {\dd \ln k}\Big{|}_{k_{\star}} = 0.968 \ee
with error bars of about $\pm 4.50\%$ for $A$ and $\pm 1.25\%$ for
$n_S$ \cite{wmap}.

In the next two sub-sections, we will focus on the quadratic
potential, $V(\phi) = (1/2) m^2\phi^2$. The equation of motion of
the scalar field (\ref{dyn}) then simplifies to:
\be \label{inflaton} \ddot\phi + 3H \dot\phi + m^2\phi =0.\ee
For this potential, $1 - n_S  = 4\epsilon $ (irrespective of the
value of the inflaton mass) so we conclude that the value of the
slow roll parameter at the time $t(k_{\star})$ is given by:
\be \label{wmap-epsilon} \epsilon (t(k_{\star})) = 8 \times 10^{-3}
\, .\ee
The observed value of the amplitude of the scalar power spectrum
then determines the Hubble parameter $H(t(k_{\star}))$ and the
Hubble radius $R_{\rm H}(t(k_{\star}))$:
\be \label{wmap-H} H(t(k_{\star})) = 7.83 \times 10^{-6}\, \mpl
\qquad{\rm or} \qquad
 R_{\rm H}(t(k_{\star})) = 1.28 \times 10^{5}\, \lp  \ee
Therefore, it follows from our discussion in section \ref{s2.1} that
\emph{the LQC corrections to general relativity are highly
suppressed at $t= t(k_{\star})$ and thereafter.} For simplicity, we
will use this fact in what follows.

Using the approximation $\epsilon= \epsilon_V$ at time
$t(k_{\star})$ (which is consistent with observational error bars)
we can determine $\phi(t(k_{\star}))$ from (\ref{wmap-epsilon}).
Then, using the implication $\epsilon/3 = \dot\phi^2/(\dot\phi^2 +
m^2\phi^2)$ of the definition of $\epsilon$, and the Friedmann
equation, it is easy to obtain the following values
\ba \phi(t(k_{\star})) = \pm 3.15\, \mpl, \qquad
\dot{\phi}(t(k_{\star})) &=& \mp 1.98\times 10^{-7} \,\mpl^2\nonumber\\
m = 1.21 \times 10^{-6}\, \mpl\, \qquad \eta(t(k_{\star})) &=&
1.61\times 10^{-2}\, . \ea
(In the reduced Planck units, often used in the cosmological
literature, the inflaton mass is $m = 6.06 \times 10^{-6} M_{\rm
pl}$.) Because this value of $m$ is somewhat different from that
given in \cite{linde-rev}, which was used in \cite{as2}, some of the
details of numerical results in this section differ from those
reported there. However, as noted in \cite{as2}, the main results
reported there do not change appreciably even if this value is
changed by a couple of orders of magnitude.

For the desired slow roll, without loss of generality we can assume
$\phi(t(k_{\star})) <0 $ and $(\dot\phi) (t(k_{\star})) >0$, so that
$\phi$ decreases as the inflaton slides down the potential. Then we
can ask for the number of e-foldings of slow roll inflation starting
from this time, $t=t(k_{\star})$ when $\epsilon = 8\times 10^{-3}$
until it increases to $0.1$. \emph{This is the `desired' slow roll
phase of interest to our analysis.} We have:
\ba \N  &:=& \ln\,\f{a_{\rm end}}{a(t(k_{\star}))} \,
=\,{\int}_{\phi(t(k_{\star}))}^{\phi_{\rm end}}\,
\f{H}{\dot\phi}\,\dd \phi\,\, = - \f{8\pi}{\mpl^2}\,
{\int}_{\phi(t(k_\star))}^{\phi_{\rm end}}\,
\f{(1+\f{\eta}{3})}{(1-\f{\epsilon}{3})}\, \f{V}{V'}\, \dd
\phi\nonumber\\
&\approx & \f{2\pi}{\mpl^2}\,
\big[(1+\f{\eta}{3})(1+\f{\epsilon}{3})\, \phi^2\big]_{\rm
end}^{t(k_{\star})}\,\, \approx
\f{2\pi}{\mpl^2}\,\big[\phi^2(t(k_{\star})) - \phi_{\rm end}^2\big]
\,\, \approx 57.5 \ea
where in the second line we have ignored second and higher order
terms in the slow roll parameters.%
\footnote{Consistency with assumptions made in the inflationary
scenario require the actual slow roll to begin somewhat earlier
because $k_{\star} = 8.58 k_o$ and the use of Bunch-Davis vacuum
requires the mode $k_o$ to be well within the Hubble radius at the
onset. The initial data we begin with ensures that this will be the
case. If we assume that at the onset the $R_{\rm H}$ is 100 times
the physical wavelength of the mode $k_o$, then there are $\sim\,
6.75$ e-foldings between the onset and $t(k_{\star})$, bringing the
total number of slow roll e-foldings from the onset until the end
(i.e. until $\epsilon =0.1$) to $\N_{\rm total}\,\sim \, 64$. This is
a lower bound; an earlier onset and hence a larger number of
e-foldings is permissible within this scenario.}
For simplicity, here we have used the general relativity field
equations. The LQC result
\be \N = \f{2\pi}{\mpl^2}\,
\big[(1+\f{\eta}{3})(1+\f{\epsilon}{3})\,\phi^2\,\,[1\,-\,
\f{V(\phi)}{2(1-\f{\epsilon}{3})\rcr}]\,\big]_{\rm
end}^{t(k_\star)}\ee
gives corrections of order $(\phi/\phi_{\rm max})^2 \, \sim \,
10^{-10}$ which are totally negligible.

To summarize, assuming the inflationary scenario with a quadratic
potential, the WMAP data provides us within observational errors of
$\lesssim\, 2\%$ : i) the value of the inflaton mass; and ii) the
initial data for inflation at some time $t(k_{\star})$ in the
early history of the universe. This data then automatically leads to
the desired slow roll inflation with $\sim\, 57$ e-foldings starting
$t=t(k_{\star})$. The question now is: What is the probability that
the LQC dynamical trajectories pass through this very small region
of the phase space, \emph{irrespective of what their initial data
are at the bounce surface?}

\subsection{Extreme kinetic energy domination at the bounce}
\label{s4.2}

In this sub-section we will study in detail the LQC dynamics of
solutions for which less than $10^{-10}$ of energy density is in the
potential at the bounce. Recall that, since the total matter density
at the bounce always equals $\rcr$, the value $|\phi_{\B}|$ of the
scalar field at the bounce point is bounded above by $\phi_{\max} =
0.90\,\mpl^2/m \approx 7.47 \times 10^{5}\,\mpl$. We will use the
fraction
\be \label{f} f:= \f{\phi_{\B}}{\phi_{\rm max}} \ee
to bin the solutions. By definition, $f$ takes values in the
interval $[-1, 1]$. Note also that $f^2 = V(\phi)/\rcr$, the
fraction of the total energy density at the bounce that is in the
potential. Therefore in this section we will focus on solutions with
$|f| <10^{-5}$. Starting at the bounce, we will first describe the
main features of the LQC evolution using suitable analytical
approximations and compare these with results obtained by numerical
evolution of the exact system.

Let us begin by noting a symmetry of the phase space equations of
motion: Given a solution $\big(\phi(t), \p(t);\, v(t), b(t)\big)$ to
(\ref{hc}) and (\ref{evo}), $\big(-\phi(t), -\p(t);\,  v(t),
b(t)\big)$ is also a solution. Therefore, in the discussion of
dynamics it suffices to focus on the initial data at the bounce
point where $\p|_{\B} \ge 0$, i.e., $\dot{\phi}_{\B} \ge 0$,
allowing $\phi_B$ to take both positive and negative values. The
numerical simulations were performed in these two sectors. The
analytical considerations presented in this section are meant to
provide a physical understanding of the main features of the early
phases of LQC dynamics. Since these do not depend on the sign of
$\phi_{\B}$, for concreteness, in the detailed discussion of
analytical considerations we will assume $\phi_{\B}$ to be
non-negative (which corresponds to numerical simulations reported in
Table I) and comment on the $\phi_{\B} <0$ case only at the end.

\subsubsection{Super-inflation}
\label{s4.2.1}

Immediately after the bounce, there is a super-inflation phase.
Since the potential energy at the bounce is very low, this phase of
evolution can be well approximated by adding only small corrections
to the analytically well-understood massless case \cite{aps3,acs}.
From the bounce to the end of super-inflation, the pair
$(H,\,\dot{H})$ goes from $(H=0,\,\,\dot{H}=10.28\, (1-f^2))$ to
$(H=0.93,\,\,\dot{H}=0)$ in Planck units. Thus, this phase is
\emph{highly} dynamical for the Hubble parameter. Let us ignore
terms of the order $f^4$. Then, the amount of time, $\Delta t$, that
super-inflation lasts is well approximated by
\be \Delta t \approx \f{\Delta H}{\dot{H}_{\avg}}\, \approx \,
\f{H_{\max}}{2\pi\dot{\phi}_{\B}^2} = \f{H_{\max}}{4\pi \rcr
(1-f^2)} \approx \f{H_{\max}}{4\pi \rcr}\, (1+f^2)\,\approx\, 0.18\,
\sp\ee
How much does the volume change? We can estimate the number of
e-foldings from the bounce to the end of super-inflation  as
follows:
\be \log(N)={\textstyle{\int}} H \dd t\, \approx\, H_{\avg} \Delta t
\, \approx \, \f{H_{\max}^2}{8 \pi \rcr}\, (1+f^2) \ee
This implies that the number of e-foldings during super-inflation is
only about $1.09$. During this time the change in the value of the
inflaton is given by:
\be \Delta \phi = {\textstyle{\int}} \dot{\phi}\, \dd t \,\approx\,
\dot{\phi}_{\avg}\, \Delta t\, \approx \,\f{1.41\, H_{\max}}{8 \pi
\sqrt{ \rcr}}\, (1-\f{f^4}{2}) \,\approx\,
0.14\,(1-\f{f^4}{2})\,\mpl \ee
which is also very small. Thus, although the change in the value of
the Hubble parameter during super-inflation is dramatic, because the
duration of this phase is so short, the total changes in the values
of the scale factor and the inflaton are also very small. While we
approximated $\dot{H}$ and $\dot\phi$ by their average values in
these calculations, the final results for $\Delta t$ and $\Delta
\phi$ are in excellent agreement with the exact numerical
calculations summarized in Table I.

\emph{Remark:} Dynamics during inflation is often described using an
analogy with a damped simple harmonic oscillator. This is because
when the slow roll conditions hold $H$ is approximately constant and
equation (\ref{inflaton}) resembles that satisfied by a damped
harmonic oscillator, with damping parameter $\zeta = 3H/2m$.
Although the universe does undergo an accelerated expansion during
super-inflation, since $H$ changes radically, intuition derived from
a damped harmonic oscillator is not useful in this phase. At the end
of super-inflation, the Hubble parameter takes its maximum value,
$H_{\max} = 0.93 \,\mpl$. Therefore, the `friction term' $\zeta =
3H/2m$ is extremely large, $\zeta \sim 1.15 \times 10^{6}$. However,
because the kinetic energy is very large
---approximately half that at the bounce--- it still
takes an appreciable time for the inflaton to slow down. During this
rather long period, dynamics is still not mimicked by a damped
harmonic oscillator. It is only when the potential energy dominates
---as is the case during the desired slow roll--- that the analogy
becomes useful.

\subsubsection{From the end of super-inflation to turn around}
\label{s4.2.2}

\begin{table}
\begin{tabular}{|c|c|c|c|c|c|c|c|}
\hline

$f$ & ${\rm Event}$ & $\phi$            & $\dot{\phi}$           & $H$                   & $\dot{H}$               & $t$                   & $\epsilon$               \\
\hline
$1.07\times 10^{-6}$ & ${\rm Bounce}$ & $8.00\times 10^{-1}$ & $ 9.05\times 10^{-1}$ & $\mathbf{0}        $ & $1.03 \times 10^{1} $ & $\mathbf{0}         $ & $\mathbf{\infty}$   \\
$                  $ & $\hbox{\rm End of SI} $ & $9.44\times 10^{-1}$ & $ 6.39\times 10^{-1}$ & $9.26\times 10^{-1}$ & $\mathbf{0}          $ & $1.80 \times 10^{-1}$ & $\mathbf{0}$        \\
$                  $ & $KE=PE       $ & $2.91              $ & $ 3.52\times 10^{-6}$ & $1.02\times 10^{-5}$ & $-1.56\times 10^{-10}$ & $4.04 \times 10^{4}$ & $ 1.50   $            \\
$                  $ & $\dot{\phi}=0$ & $3.03              $ & $ \mathbf{0}        $ & $7.50\times 10^{-6}$ & $\mathbf{0}          $ & $1.66 \times 10^{5}$ & $\mathbf{0}$         \\
$                  $ & $\epsilon=8 \times 10^{-3}$ & $3.01 $ & $-1.88\times 10^{-7}$ & $7.46\times 10^{-6}$ & $-4.45\times 10^{-13}$ & $3.05 \times 10^{5}$ & $8\times 10^{-3}$ \\
\hline
$1.17\times 10^{-6}$ & ${\rm Bounce}$ & $8.75\times 10^{-1}$ & $ 9.05\times 10^{-1}$ & $\mathbf{0}        $ & $1.03 \times 10^{1} $ & $\mathbf{0}         $ & $\mathbf{\infty} $\\
$                  $ & $\hbox{\rm End of SI}  $ & $1.02              $ & $ 6.39\times 10^{-1}$ & $9.26\times 10^{-1}$ & $\mathbf{0}          $ & $1.80 \times 10^{-1}$ & $\mathbf{0}       $  \\
$                  $ & $KE=PE       $ & $2.98              $ & $ 3.61\times 10^{-6}$ & $1.04\times 10^{-5}$ & $-1.63\times 10^{-10}$ & $3.94 \times 10^{4}$ & $1.50  $\\
$                  $ & $\dot{\phi}=0$ & $3.10              $ & $ \mathbf{0}        $ & $7.67\times 10^{-6}$ & $\mathbf{0}          $ & $1.63 \times 10^{5}$ & $\mathbf{0}  $       \\
$                  $ & $\epsilon=8 \times 10^{-3}$ & $3.07 $ & $ 1.92\times 10^{-7}$ & $7.62\times 10^{-6}$ & $-4.56\times 10^{-13}$ & $3.28 \times 10^{5}$ & $8\times 10^{-3}$ \\
\hline
$1.25\times 10^{-6}$ & ${\rm Bounce}$ & $9.37\times 10^{-1}$ & $ 9.05\times 10^{-1}$ & $\mathbf{0}        $ & $1.03 \times 10^{1} $ & $\mathbf{0}         $ & $\mathbf{\infty}$ \\
$                  $ & $\hbox{\rm End of SI}  $ & $1.08              $ & $ 6.39\times 10^{-1}$ & $9.26\times 10^{-1}$ & $\mathbf{0}          $ & $1.80 \times 10^{-1}$ & $\mathbf{0}      $   \\
$                  $ & $KE=PE       $ & $3.04              $ & $ 3.67\times 10^{-6}$ & $1.06\times 10^{-5}$ & $-1.69\times 10^{-10}$ & $3.87 \times 10^{4}$ & $1.50  $\\
$                  $ & $\dot{\phi}=0$ & $3.16              $ & $ \mathbf{0}        $ & $7.82\times 10^{-6}$ & $\mathbf{0}          $ & $1.60 \times 10^{5}$ & $\mathbf{0} $        \\
$                  $ & $\epsilon=8 \times 10^{-3}$ & $3.12 $ & $-1.95\times 10^{-7}$ & $7.75\times 10^{-6}$ & $-4.81\times 10^{-13}$ & $3.75 \times 10^{5}$ & $8\times 10^{-3}$\\
\hline
$1.42\times 10^{-6}$ & ${\rm Bounce}$ & $1.06              $ & $ 9.05\times 10^{-1}$ & $\mathbf{0}        $ & $1.03 \times 10^{1} $ & $\mathbf{0}         $ & $\mathbf{\infty} $\\
$                  $ & $\hbox{\rm End of SI}  $ & $1.20              $ & $ 6.39\times 10^{-1}$ & $9.26\times 10^{-1}$ & $\mathbf{0}          $ & $1.80 \times 10^{-1}$ & $\mathbf{0}       $  \\
$                  $ & $KE=PE       $ & $3.16              $ & $ 3.81\times 10^{-6}$ & $1.10\times 10^{-5}$ & $-1.82\times 10^{-10}$ & $3.73 \times 10^{4}$ & $1.50  $\\
$                  $ & $\dot{\phi}=0$ & $3.28              $ & $ \mathbf{0}        $ & $8.11\times 10^{-6}$ & $\mathbf{0}          $ & $1.56 \times 10^{5}$ & $\mathbf{0}$         \\
$                  $ & $\epsilon=8 \times 10^{-3}$ & $3.15 $ & $-1.97\times 10^{-7}$ & $7.82\times 10^{-6}$ & $-4.86\times 10^{-13}$ & $8.23 \times 10^{5}$ & $8\times 10^{-3}$ \\
\hline
$1.59\times 10^{-6}$ & ${\rm Bounce}$ & $1.19              $ & $ 9.05\times 10^{-1}$ & $\mathbf{0}        $ & $1.03 \times 10^{1} $ & $\mathbf{0}         $ & $\mathbf{\infty} $\\
$                  $ & $\hbox{\rm End of SI}  $ & $1.33              $ & $ 6.39\times 10^{-1}$ & $9.26\times 10^{-1}$ & $\mathbf{0}          $ & $1.80 \times 10^{-1}$ & $\mathbf{0}        $ \\
$                  $ & $KE=PE       $ & $3.28              $ & $ 3.96\times 10^{-6}$ & $1.15\times 10^{-5}$ & $-1.97\times 10^{-10}$ & $3.59 \times 10^{4}$ & $1.50  $\\
$                  $ & $\dot{\phi}=0$ & $3.40              $ & $ \mathbf{0}        $ & $8.42\times 10^{-6}$ & $\mathbf{0}          $ & $1.51 \times 10^{5}$ & $\mathbf{0}         $\\
$                  $ & $\epsilon=8 \times 10^{-3}$ & $3.15 $ & $-1.97\times 10^{-7}$ & $7.82\times 10^{-6}$ & $-4.87\times 10^{-13}$ & $1.45 \times 10^{6}$ & $8\times 10^{-3} $\\
\hline
$1.77\times 10^{-6}$ & ${\rm Bounce}$ & $1.32              $ & $ 9.05\times 10^{-1}$ & $\mathbf{0}        $ & $1.03 \times 10^{1} $ & $\mathbf{0}         $ & $\mathbf{\infty} $\\
$                  $ & $\hbox{\rm End of SI}  $ & $1.46              $ & $ 6.39\times 10^{-1}$ & $9.26\times 10^{-1}$ & $\mathbf{0}          $ & $1.80 \times 10^{-1}$ & $\mathbf{0}      $   \\
$                  $ & $KE=PE       $ & $3.40              $ & $ 4.11\times 10^{-6}$ & $1.19\times 10^{-5}$ & $-2.13\times 10^{-10}$ & $3.46 \times 10^{4}$ & $1.50  $\\
$                  $ & $\dot{\phi}=0$ & $3.52              $ & $ \mathbf{0}        $ & $8.73\times 10^{-6}$ & $\mathbf{0}          $ & $1.47 \times 10^{5}$ & $\mathbf{0} $        \\
$                  $ & $\epsilon=8 \times 10^{-3}$ & $3.15 $ & $-1.97\times 10^{-7}$ & $7.82\times 10^{-6}$ & $-4.87\times 10^{-13}$ & $2.08 \times 10^{6}$ & $8\times 10^{-3} $\\
\hline
\end{tabular}
\caption{Dynamical evolution of $\phi$ and $H$ for various values of
$f$ in the case $\phi_{\B} >0$. Analytical considerations imply that
certain entries should be identically zero; their values are are
$O(\epsilon_{machine})$ in simulations but denoted by a boldface
zero ($\mathbf{0}$) in the Table. Events considered are: the bounce
point, end of super-inflation (End of SI), equality of potential and
kinetic energy ($KE=PE$), turn around ($\dot{\phi}=0$), and reaching
$\epsilon=8\times 10^{-3}$. The desired slow roll compatible with
the WMAP data is not realized if $f \le 1.17 \times 10^{-6}$ but is
realized for $f \ge 1.25 \times 10^{-6}$.}
\end{table}

Let us now consider the post-super-inflation phase. An analytical
calculation can be carried out in two steps with approximations
tailored to each case: i) the phase which commences at the end of
super-inflation and ends when the kinetic energy equals the
potential energy; and ii) the phase between the time this equality
is reached and turn-around of the inflaton, i.e., when the kinetic
energy reduces to zero. Dynamics has qualitatively different
features in these two phases because the first is dominated by
kinetic energy and the second by potential.

In phase i), the ratio of the potential to the kinetic energy is
function $\alpha(t)$ which takes values in $(0,\, 1)$. The
approximation consists of replacing it with its `average' value
$\alpha_o$. At the end of the calculation, $\alpha_o$ is determined
by comparing the analytic expression with the exact numerical answer
in one of the cases given in Table I and then used in other cases.
We start with the equation $\dot{b} = -4\pi\gamma {\dot\phi}^2$ from
(\ref{evo}) and, using the constraint equation (\ref{hc}) and the
approximation, obtain a differential equation containing only $b$.
The solution $b(t)$ is given by
\be \label{b} \cot \lambda b(t) - \cot \lambda b(t_e)  = \f{2.4439\,
\lambda\, (t-t_e)}{1+\alpha_o}\,  \ee
where $t_e$ is the proper time at the end of super-inflation. It can
be substituted back in the original Friedmann equation to obtain
$t-t_e$ as a function of $\phi(t)-\phi(t_e)$.  Finally, we set $t=
t_{\rm Eq}$, the time at which kinetic and potential energies are
equal, to obtain:
\be \phi(t_{\rm Eq}) - \phi(t_e) = 0.1628\, \sqrt{1+\alpha_o}\,
(y_{\rm Eq} - y_e)\, \quad {\rm where} \quad  {\rm sinh}(y) = 1 +
(\f{5.56}{1+\alpha_o}) t \, .\ee
Next, the constraint equation (\ref{hc}) determines $\phi(t_{\rm
Eq})$ in terms of $b(t_{\rm Eq})$ and we can invert the expression
(\ref{b}) of $b(t)$ to eliminate it in favor of $t_{\rm Eq} -t_e$.
The result is an expression relating $t_{\rm Eq} - t_e$ and
$\phi(t_e)$ alone. It determines how long the first
post-super-inflation phase lasts for any given value $\phi_e$ of the
inflaton at the end of super-inflation:
\ba \label{phie} \phi_e &=& \f{1}{m}\,
\f{0.640}{\sqrt{1+\big(1+\f{5.56}{1+\alpha_o}\, (t_{\rm
Eq}-t_e)\big)^2}}\, -\, 0.1628\,\sqrt{1+\alpha_o}\, (y_{\rm
Eq} - y_e)\nonumber\\
&\approx& \f{9.508\times 10^4\, (1+\alpha_o)\,}{t_{\rm Eq}}\, -\,
0.1628\sqrt{1+\alpha_o}\, \big( {\rm sinh}^{-1}\,(1+\f{5.56 (t_{\rm
Eq})}{1+\alpha_o})\,-\, 0.88\big) \, \ea
where for simplicity we set $t_e=0$ in the last step. Comparison of
this expression with the entry $f= 1.77\times 10^{-6}$ in table
Table I, one finds that the fit with the exact numerical result is
best when $\alpha_o =1/3$. We will use this value and compare the
approximate analytical expression with the exact numerical result in
all cases covered in Table I. The resulting Table II shows that the
agreement is very good; the analytical approximation we made by
replacing $\alpha$ with its `average' $\alpha_o$ works quite well.
Table II also shows that the duration $t_{\rm Eq}-t_e \sim
10^{4}\,\sp$ of phase i) is \emph{much} longer than that of
super-inflation.

\begin{table}
\begin{tabular}{|c|c|c|c|c|c|c|}
\hline

${\rm Method}$ & ${\phi_e=9.44\times 10^{-1}}$ & $\phi_e=1.02$ & $\phi_e=1.08$ & $\phi_e=1.20$ & $\phi_e=1.33$ & $\phi_e=1.46$  \\
\hline
${\hbox{\rm Exact Numerics}}$ & $ 4.04\times 10^4$ &  $ 3.94\times 10^4$ & $3.87\times 10^4$ & $3.73\times 10^4$ & $3.59\times 10^4$ & $3.46\times 10^4$   \\
\hline
${\hbox{\rm Analytic Eq}} (\ref{phie})$ & $ 4.00\times 10^4$ & $3.91\times 10^4$ & $3.84\times 10^4$ & $3.71\times 10^4$ & $3.58\times 10^4$ & $3.46\times10^4$ \\
\hline
\end{tabular}
\caption{The time interval $t_{\rm Eq}-t_e$ as a function of the
value $\phi_e$ of in the inflaton at the end of super-inflation:
Comparison between the exact numerical result and the analytical
approximation. As $\phi_e$ increases, the difference between kinetic
and potential energies at the end of super-inflation decreases,
whence it less takes less time to achieve equality between potential
and kinetic energy. The analytical approximation replaced the ratio
$\alpha(t)$ between the two energies by an `average' $\alpha_o$
during this phase i) after the end of super-inflation.}
\end{table} \bigskip

Eq (\ref{phie}) determines the duration $t_{\rm Eq} -t_e$ of phase
i) as a function of the value of the inflation at the end of
super-inflation. The same approximation as was used there provides
expressions for values of $H,\phi,\dot\phi$ at the end of phase i)
as a function of $t_{\rm Eq}-t_e$:
\be \label{Eq} H(t_{\rm Eq}) \approx \f{1+\alpha_o}{3.00 (t_{\rm Eq}
- t_e)};\quad \dot\phi(t_{\rm Eq}) \approx
\f{0.1152(1+\alpha_o)}{t_{\rm Eq}-t_e};\quad \phi(t_{\rm Eq})\approx
\f{9.5207\times 10^4 \, (1+\alpha_o)}{t_{\rm Eq}-t_e}\, . \ee
Again, there is good agreement with the exact numerical results. For
the Hubble parameter a comparison is given in Table III.

\bigskip
\begin{table}
\begin{tabular}{|c|c|c|c|c|c|c|}
\hline
${\rm Method}$ & ${\phi_e=9.44\times 10^{-1}}$ & $\phi_e=1.02$ & $\phi_e=1.08$ & $\phi_e=1.20$ & $\phi_e=1.33$ & $\phi_e=1.46$  \\
\hline
${\hbox{\rm Exact Numerics}}$ & $ 1.02\times 10^{-5}$ &  $ 1.04\times 10^{-5}$ & $1.06\times 10^{-5}$ & $1.10\times 10^{-5}$ & $1.15\times 10^{-5}$ & $1.19\times 10^{-5}$   \\
\hline
${\hbox{\rm Analytic Eq}} (\ref{Eq})$ & $ 1.11\times 10^{-5}$ & $1.13\times 10^{-5}$ & $1.15\times10^{-5}$ & $1.19\times 10^{-5}$ & $1.24\times 10^{-5} $ & $1.28\times 10^{-5}$ \\
\hline
\end{tabular}
\caption{Value of the Hubble parameter $H$ at $t=t_{\rm Eq}$ as a
function of the value $\phi_e$: Comparison between the exact
numerical result and the analytical approximation. The analytical
approximation replaced the ratio $\alpha(t)$ between the two
energies by an `average' $\alpha_o$ during this phase i) after the
end of super-inflation. The resulting error is less than 10\%. Since
$\phi_e \approx \phi_{\B} = f \phi{\max}$,\, $H(t_{\rm Eq})$
increases monotonically with $f$. Since $H(t_{\rm Eq}) \approx
10^{-5}\mpl$, general relativity is an excellent approximation to
LQC at $t=t_{\rm Eq}$.}
\end{table}
%

Thus the analytic approximation yields a good portrait of dynamics
of all physical quantities of interest. Note that the Hubble
parameter changes very significantly during this phase also: At the
end of super-inflation, we have $H \approx 0.93\, \mpl$ while at
$t_{\rm Eq}$, $H \approx 10^{-5}\mpl$. Similarly, while $\dot\phi
\approx 0.9 \mpl^2$ at the end of super-inflation, we have
$\dot\phi(t_{\rm Eq}) \approx 4 \times 10^{-6} \mpl^2$.

Let us now turn to the subsequent phase ii) which lasts until the
inflaton turns around, i.e., its kinetic energy vanishes
identically. Since the energy density has dramatically decreased
during phase i)  ---at $t_{\rm Eq}$, the matter density $\rho(t_{\rm
Eq})$ is of the order $10^{-11} \rp$ in all cases given in Table
III--- general relativity is an excellent approximation in this
phase. Also, since the kinetic energy is sub-dominant, in contrast
to phase i), phase ii) is dynamically rather quiescent. At the start
of this phase, values of $H, \phi, \dot\phi$ are given by
(\ref{Eq}). Using them and (\ref{inflaton}) we can calculate
$\ddot\phi (t_{\rm Eq})$ and $\dddot\phi(t_{\rm Eq})$. In all cases
summarized in Table III, $\ddot\phi (t_{\rm Eq}) \sim 10^{-10}
\mpl^2$ and $\dddot\phi(t_{\rm Eq}) \lesssim 10^{-16} \mpl^3$.
Because of this quiescent behavior we can hope to obtain the time
duration $(t_{\rm TA} - t_{\rm Eq})$ until the turn around by a
simple calculation:
\ba \label{duration} (t_{\rm TA} - t_{\rm Eq}) &\approx& \f{\Delta
\dot\phi}{\ddot\phi_{\avg}} \approx - \f{\dot\phi
(t_{\rm Eq})}{\beta\ddot\phi(t_{\rm Eq})}\nonumber\\
&\approx& \f{\dot\phi(t_{\rm Eq})}{\beta(3H(t_{\rm
Eq})+m)\dot\phi(t_{\rm Eq})} = \f{1}{\beta(3H(t_{\rm Eq})+m)}\ea
where $\beta$ is a coefficient (of $O(1)$) relating the value of
$\ddot\phi$ at $t= t_{\rm Eq}$ to its average value in phase ii).
(As explained in the caption of Table IV, $\beta$ was set to $1/4$
in numerical evaluations of (\ref{duration}).)
\bigskip
\begin{table}
\begin{tabular}{|c|c|c|c|c|c|c|}
\hline
${\rm Method}$ & ${\phi_e=9.44\times 10^{-1}}$ & $\phi_e=1.02$ & $\phi_e=1.08$ & $\phi_e=1.20$ & $\phi_e=1.33$ & $\phi_e=1.46$  \\
\hline
${\hbox{\rm Exact Numerics}}$ & $ 1.26\times 10^{5}$ &  $ 1.24\times 10^{5}$ & $1.21\times 10^{5}$ & $1.19\times 10^{5}$ & $1.15\times 10^{5}$ & $1.12\times 10^{5}$   \\
\hline
${\hbox{\rm Analytic Eq}} (\ref{Eq})$ & $ 1.26\times 10^{5}$ & $1.23\times 10^{5}$ & $1.21\times10^{5}$ & $1.17\times 10^{5}$ & $1.12\times 10^{5} $ & $1.08\times 10^{5}$ \\
\hline
\end{tabular}
\caption{Duration of phase ii), $t_{\rm TA} -t_{\rm Eq}$, as a
function of $\phi_e$: Comparison between the exact numerical result
and the analytical approximation. The analytical approximation
assumes that $\ddot\phi_{\rm Avg}$ is given by $\beta\,
\times\,\ddot\phi(t_{\rm Eq})$ where $\beta$ is set equal to $1/4$
by agreement with numerical values for $\phi_e= 0.944\mpl$. the
analytical expression then reproduces the exact numerical results
for other values of $\phi_e$ to within 4\%.}
\end{table}
%

Table IV provides a comparison with the exact numerical values and
those calculated using (\ref{duration}). Note that the quiescent
phase ii) lasts even longer than the more dynamic phase i). In the
case of super-inflation we could use the `averaging approximation'
because, although the Hubble parameter is very dynamic, the phase is
very short lived. In phase ii) we could use it because although the
phase is long lived, it is quiescent. Finally, although phase ii)
lasts $\sim 10^5 \, \sp$, because $\dot{H},\, \dot\phi$ are
$\lesssim 10^{-10},\, \lesssim 10^{-6}$ at $t=t_{\rm Eq}$, the
change in $H,\,\phi$ during phase ii) is quite small, again
highlighting the quiescent nature of the phase.

\begin{figure}[ht]  \includegraphics [width=6in] {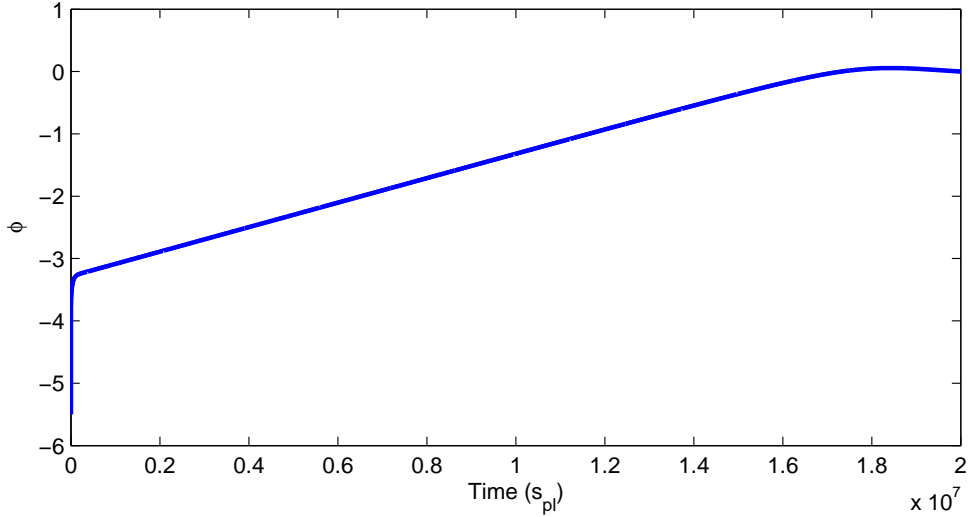}
\caption{ Illustrative evolution of the scalar field. In this plot we
use a negative value of $\phi_{\B}$ to complement the analytical discussion:
\, $\phi_{\B} = -5.5\, \mpl$ and $\dot{\phi}_{\B} >0$. (See the case
$f= -7.35\times 10^{-6}$ in Table V.) The inflaton rolls down the
potential quickly at first because the kinetic energy at the bounce
is large. The kinetic energy equals potential energy at $t=t_{\rm Eq}
\equiv 3.51 \times 10^{4}\, \sp$ and $\phi$ changes more slowly during
the subsequent quiescent phase. The WMAP value $\epsilon= 8\times
10^{-3}$ is reached $t= 7.08 \times 10^{5}\,\sp$ and the end of slow roll,
$\epsilon=0.1$, used in this paper is reached at $t= 1.24\times 10^7\, \sp$.
Because $\phi_{\B}$ is negative, in this case the inflaton turns
around after the end of the desired slow roll, at $t=1.84\times 10^{7}\,
\sp$.} \label{Fig2}
\end{figure}
\begin{figure}[ht] \includegraphics  [width=6in] {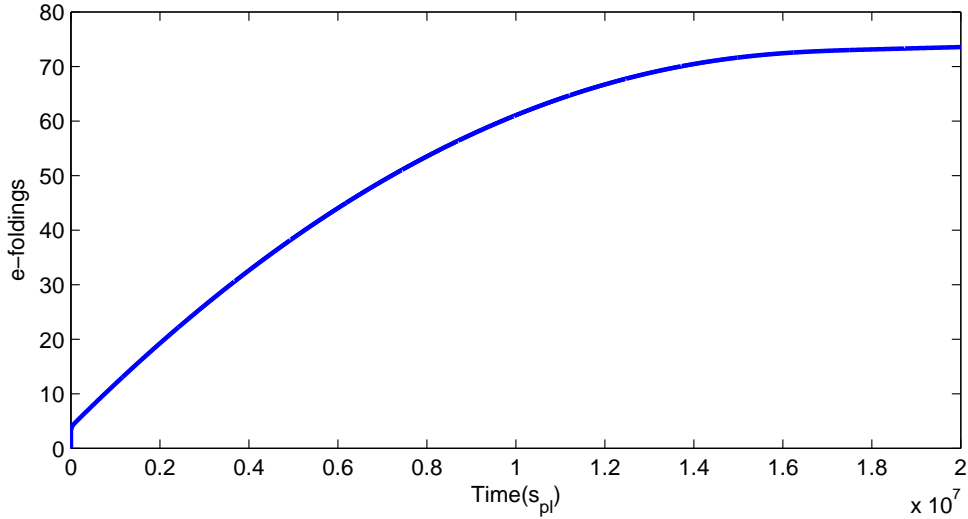}
\caption{Illustrative evolution of the scale factor for the same case
as in Fig \ref{Fig2}, i.e., for $\phi_{\B} = -5.5\, \mpl$ and
$\dot{\phi}_{\B} >0$. (See the case $f= -7.35\times 10^{-6}$ in Table V.)
The WMAP value $\epsilon= 8\times 10^{-3}$ is reached $t= 7.08 \times
10^{5}\,\sp$ and the end of slow roll, $\epsilon=0.1$, used in this paper
is reached at $t= 1.24\times 10^7\, \sp$. The number of e-foldings flattens
out soon after this time.}
\label{Fig3} \end{figure}

\subsubsection{Post turn-around}
\label{s4.2.3}

At the end of phase ii), the inflaton turns around and starts
rolling down the potential. Since by definition $\dot\phi (t_{\rm
TA}) =0$, the slow roll parameter $\epsilon$ also vanishes there.
However, because $\dot{H} \propto \dot\phi^2$, it also vanishes
there, whence the slow roll parameter $\eta$ need not be small and
we are not assured the beginning of a slow roll. The desired slow
roll, if it is realized in the given solution, must occur
subsequently, as the inflaton rolls down the potential.

Recall that, for the desired slow roll to be realized, we must have
$\phi(t(k_{\star})) \approx 3.15\, \mpl$. This point can be reached
if and only if $\phi(t_{\rm TA}) > 3.15\,\mpl$. When is this
condition realized and when does it fail to be realized in the
extreme kinetic energy dominated case under consideration? The
analytical formulas of the last two sub-sections show that $\phi
(t_{\rm TA})$ increases with $f$ and the illustrative numerical
values we have presented ensure that $\phi(t_{\rm TA}) > 3.15\,
\mpl$ if $\phi_{\B} \ge 0.937\, \mpl$.
(This is a sufficient condition rather than a sharp bound.) Since
$|\phi|\,\in (0,\, 7.47\,\mpl)$ in the extreme kinetic energy
dominated case, our result suggests that it is likely that the
desired conditions are met. However, in the analytical
considerations we focused only on the $\phi_{\B} \ge0$ case and
moreover made some approximations. Therefore, we also carried out a
large number of numerical simulations of exact equations to first
confirm and then sharpen this implication. (Tables I and V contain
only a few illustrative results.)

The simulations were performed in MATLAB using a Runge-Kutta (4,5)
ordinary differential equation solver (ODE45) with both relative and
absolute tolerances set at $3 \times 10^{-14}$. To facilitate
computation, the logarithm of the ratio of the volume at a given
time to the volume at the bounce was chosen to be a fundamental
variable. On each solution the preservation of the Hamiltonian
constraint was verified at the level of the tolerances. For
robustness sample solutions were also evolved independently with
various different tolerances, using a set of equations which did not
involve the initial volume, and seen to agree with those given by
the Runge-Kutta algorithm. Time scales for the evolutions were
chosen such that the numerical noise introduced by the deviation of
the Hamiltonian constraint from zero would not contribute
significantly (i.e. greater than one part in $10^6$) to the
fundamental variables. Separate searches were used around the end
points of the allowed space, i.e., very low values of $f$ for which
the dynamical trajectories fail to be consistent with the WMAP data.
This ensured that the extremely high probability we found for the
LQC dynamics to be compatible with the WMAP data is in fact a lower
bound.

\begin{table}
\begin{tabular}{|c|c|c|c|c|c|c|c|}
\hline

$f$ & ${\rm Event}$ & $\phi$            & $\dot{\phi}$           & $H$                   & $\dot{H}$               & $t$                   & $\epsilon$               \\
\hline $-7.02\times 10^{-6}$& ${\rm Bounce}$ & $-5.25 $ &
$9.05\times 10^{-1}$ & $\mathbf{0}       $ & $1.03 \times 10^{1}$ &
$\mathbf{0} $ & $\mathbf{\infty}$ \\ $
                   $ & $\hbox{\rm End of SI}$ & $-5.11 $ & $6.39\times 10^{-1}$ & $9.26\times 10^{-1}$ & $\mathbf{0}       $ & $1.80\times 10^{-1}$ & $\mathbf{0}       $ \\ $
                   $ & $KE=PE$ & $-3.15 $ & $3.80\times 10^{-6}$ & $1.10\times 10^{-5}$ & $-1.82\times 10^{-10}$ & $3.81 \times 10^{4}$ & $1.50 $ \\ $
                   $ & $\dot{\phi}=0$ & $5.42\times 10^{-2}$ & $\mathbf{0}       $ & $2.25\times 10^{-7}$ & $\mathbf{0}       $ & $1.71 \times 10^{7}$ & $\mathbf{0}       $ \\ $
                   $ & $\epsilon=8 \times 10^{-3}$ & $NA$ & $NA$ & $NA$ & $NA$ & $NA$ & $NA$ \\
\hline $-7.22\times 10^{-6} $ & ${\rm Bounce}$ & $-5.40 $ &
$9.05\times 10^{-1}$ & $\mathbf{0}       $ & $1.03 \times 10^{1}$ &
$\mathbf{0} $ & $\mathbf{\infty}$ \\ $
                   $ & $\hbox{\rm End of SI}$ & $-5.26 $ & $6.39\times 10^{-1}$ & $9.26\times 10^{-1}$ & $\mathbf{0}       $ & $1.80\times 10^{-1}$ & $\mathbf{0}       $ \\ $
                   $ & $KE=PE$ & $-3.31 $ & $4.00\times 10^{-6}$ & $1.16\times 10^{-5}$ & $-2.01\times 10^{-10}$ & $3.62 \times 10^{4}$ & $1.50 $ \\ $
                   $ & $\dot{\phi}=0$ & $5.64\times 10^{-2}$ & $\mathbf{0}       $ & $2.14\times 10^{-7}$ & $\mathbf{0}       $ & $1.79 \times 10^{7}$ & $\mathbf{0}       $ \\ $
                   $ & $\epsilon=8 \times 10^{-3}$ & $NA$ & $NA$ & $NA$ & $NA$ & $NA$ & $NA$ \\
\hline $-7.35\times 10^{-6} $ & ${\rm Bounce}$ & $-5.50 $ &
$9.05\times 10^{-1}$ & $\mathbf{0}       $ & $1.03 \times 10^{1}$ &
$\mathbf{0} $ & $\mathbf{\infty}$ \\ $
                   $ & $\hbox{\rm End of SI}$ & $-5.36 $ & $6.39\times 10^{-1}$ & $9.26\times 10^{-1}$ & $\mathbf{0}       $ & $1.80\times 10^{-1}$ & $\mathbf{0}       $ \\ $
                   $ & $KE=PE$ & $-3.41 $ & $4.12\times 10^{-6}$ & $1.19\times 10^{-5}$ & $-2.14\times 10^{-10}$ & $3.51 \times 10^{4}$ & $1.50 $ \\ $
                   $ & $\dot{\phi}=0$ & $5.51\times 10^{-2}$ & $\mathbf{0}       $ & $2.21\times 10^{-7}$ & $\mathbf{0}       $ & $1.84 \times 10^{7}$ & $\mathbf{0}       $ \\ $
                   $ & $\epsilon=8 \times 10^{-3}$ & $-3.14 $ & $1.97\times 10^{-7}$ & $7.80\times 10^{-6}$ & $-4.86\times 10^{-13}$ & $7.08 \times 10^{5}$ & $8 \times 10^{-3}$ \\
\hline $-7.69\times 10^{-6} $ & ${\rm Bounce}$ & $-5.75 $ &
$9.05\times 10^{-1}$ & $\mathbf{0}       $ & $1.03 \times 10^{1}$ &
$\mathbf{0} $ & $\mathbf{\infty}$ \\ $
                   $ & $\hbox{\rm End of SI}$ & $-5.61 $ & $6.39\times 10^{-1}$ & $9.26\times 10^{-1}$ & $\mathbf{0}       $ & $1.80\times 10^{-1}$ & $\mathbf{0}       $ \\ $
                   $ & $KE=PE$ & $-3.68 $ & $4.44\times 10^{-6}$ & $1.29\times 10^{-5}$ & $-2.48\times 10^{-10}$ & $3.26 \times 10^{4}$ & $1.50 $ \\ $
                   $ & $\dot{\phi}=0$ & $5.71\times 10^{-2}$ & $\mathbf{0}       $ & $2.11\times 10^{-7}$ & $\mathbf{0}       $ & $1.97 \times 10^{7}$ & $\mathbf{0}       $ \\ $
                   $ & $\epsilon=8 \times 10^{-3}$ & $-3.14 $ & $1.97\times 10^{-7}$ & $7.80\times 10^{-6}$ & $-4.87\times 10^{-13}$ & $2.03 \times 10^{6}$ & $8 \times 10^{-3}$ \\
\hline $-7.92\times 10^{-6} $ & ${\rm Bounce}$ & $-5.92 $ &
$9.05\times 10^{-1}$ & $\mathbf{0}       $ & $1.03 \times 10^{1}$ &
$\mathbf{0} $ & $\mathbf{\infty}$ \\ $
                   $ & $\hbox{\rm End of SI}$ & $-5.78 $ & $6.39\times 10^{-1}$ & $9.26\times 10^{-1}$ & $\mathbf{0}       $ & $1.80\times 10^{-1}$ & $\mathbf{0}       $ \\ $
                   $ & $KE=PE$ & $-3.85 $ & $4.66\times 10^{-6}$ & $1.35\times 10^{-5}$ & $-2.73\times 10^{-10}$ & $3.10 \times 10^{4}$ & $1.50 $ \\ $
                   $ & $\dot{\phi}=0$ & $5.48\times 10^{-2}$ & $\mathbf{0}       $ & $2.22\times 10^{-7}$ & $\mathbf{0}       $ & $2.07 \times 10^{7}$ & $\mathbf{0}       $ \\ $
                   $ & $\epsilon=8 \times 10^{-3}$ & $-3.14 $ & $1.97\times 10^{-7}$ & $7.80\times 10^{-6}$ & $-4.86\times 10^{-13}$ & $2.94 \times 10^{6}$ & $8 \times 10^{-3}$ \\
\hline $-8.02\times 10^{-6} $ & ${\rm Bounce}$ & $-6.00 $ &
$9.05\times 10^{-1}$ & $\mathbf{0}       $ & $1.03 \times 10^{1}$ &
$\mathbf{0} $ & $\mathbf{\infty}$ \\ $
                   $ & $\hbox{\rm End of SI}$ & $-5.86 $ & $6.39\times 10^{-1}$ & $9.26\times 10^{-1}$ & $\mathbf{0}       $ & $1.80\times 10^{-1}$ & $\mathbf{0}       $ \\ $
                   $ & $KE=PE$ & $-3.94 $ & $4.75\times 10^{-6}$ & $1.38\times 10^{-5}$ & $-2.84\times 10^{-10}$ & $3.04 \times 10^{4}$ & $1.50 $ \\ $
                   $ & $\dot{\phi}=0$ & $5.51\times 10^{-2}$ & $\mathbf{0}       $ & $2.21\times 10^{-7}$ & $\mathbf{0}       $ & $2.11 \times 10^{7}$ & $\mathbf{0}       $ \\ $
                   $ & $\epsilon=8 \times 10^{-3}$ & $-3.14 $ & $1.97\times 10^{-7}$ & $7.80\times 10^{-6}$ & $-4.86\times 10^{-13}$ & $3.36 \times 10^{6}$ & $8 \times 10^{-3}$ \\

\hline
\label{table1}
\end{tabular}

\caption{Dynamical evolution of $\phi$ and $H$ for various values of
$f$ in the case $\phi_{\B} <0$. Analytical considerations imply that
certain entries should be identically zero; their values are are
$O(\epsilon_{machine})$ in simulations but denoted by a boldface
zero ($\mathbf{0}$) in the Table. Events considered are: the bounce
point, end of super-inflation (End of SI), equality of potential and
kinetic energy ($KE=PE$), turn around ($\dot{\phi}=0$), and reaching
$\epsilon=8\times 10^{-3}$. Because $\phi_{\B} <0$, the turn around
happens \emph{after} the desired inflation ends. The desired slow
roll compatible with the WMAP data is not realized if $|f| \le 7.22
\times 10^{-6}$ but is realized if $|f| \ge 7.35 \times 10^{-6}$.}
\end{table}

As discussed in section \ref{s4.1}, for the quadratic potential the
WMAP data implies that there was a time $t(k_\star)$ in the early
history of the universe when $\epsilon= 8\times 10^{-3}$ and $\phi=
3.15\,\mpl$  within error bars (of $\lesssim 4.5\%$). In LQC we can
specify the initial data at the bounce. Numerical simulations let us
answer the following question for the extreme kinetic energy
dominated bounce: What are restrictions on the initial data at the
bounce for the ensuing LQC dynamical trajectory to meet the WMAP
constraint? Full numerical simulations bear out conclusions
suggested by Tables I and V:
\be \label{suff}
 {\hbox{\rm WMAP constraint is met provided}}\,\,\,
\begin{cases}
f \ge 1.25 \times 10^{-6} &  \text{if $\phi_{\B} \ge 0$}\\
|f| \ge 7.35 \times 10^{-6} &\text{if $\phi_{\B} <0$\, .}
\end{cases}\, \ee
Thus, in the extreme kinetic dominated case, \emph{we are guaranteed
that the event $E$ ---the desired slow roll compatible with WMAP
observations--- will be realized in the solution under consideration
if $f$ is outside the interval $(-7.35 \times 10^{-6},\, 1.25 \times
10^{-6})$.} (Again, this is only a sufficient condition.) Since full
range of $f$ in the kinetic energy dominated case under
consideration is $(-10^{-5},\, 10^{-5})$, this result suggests that
$E$ will be realized in `more than half the trajectories' in this
case. We will sharpen this statement using the normalized Liouville
measure in section \ref{s4.4}.

The precise numbers in the sufficient condition (\ref{suff}) depend
on the sign of $\phi_{\rm B}$ at the bounce. We will conclude with a
brief discussion of the origin of the difference between the
$\phi_{\B}>0$ and $\phi_{\B} <0$ cases. The origin of the asymmetry
lies in the fact that (because of symmetries of field equations) we
have restricted ourselves to data with $\dot{\phi}_{\B} \ge 0$. So,
if $\phi_{\B}<0$, the inflaton is already rolling down the potential
at the bounce. If initially it is sufficiently high up in the
potential with, $|\phi_{\B}| > 5.5\,\mpl$, friction can slow it down
sufficiently for the desired slow roll to begin at $\phi \approx
-3.15\,\mpl$. But if $|\phi_{\B}| < 5.5\,\mpl$, then the kinetic
energy is too large for the friction term to slow it sufficiently
for the desired slow roll to commence before it reaches the bottom
of the potential. Then it starts climbing up on the $\phi
>0$ branch but does not acquire a value higher that $\sim 3.15$
required for the dynamical trajectory to pass through the region
satisfying the WMAP constraints. In the $\phi_{\B}>0$ case, by
contrast, if $\phi_{\B} \gtrsim 0.94$, the kinetic energy at the
bounce is sufficient to propel the inflaton to values higher than
$3.15\mpl$ on the $\phi_{\B} >0$ branch of the potential to allow
for the desired slow roll.

In subsection \ref{s4.3.2}, we will show that the likelihood of
attaining the desired slow roll is in fact $1$ if the bounce is
\emph{not} extreme kinetic energy dominated and in section
\ref{s4.4} we will calculate the precise probability for the
occurrence of $E$ on the entire space $\S$ of solutions using the
normalized Liouville measure.

\subsection{Generic LQC solutions and the WMAP slow roll}
\label{s4.3}

In section \ref{s4.3.1} we continue our description of
qualitative features of LQC dynamics, now for bounces with
$f>10^{-5}$. This discussion is rather sketchy compared to that
of section \ref{s4.2} because in this case there are no
subtleties with respect to the WMAP data: as we show in section
\ref{s4.3.2}, all dynamical trajectories meeting this condition
at the bounce satisfy the WMAP  constraints.

\subsubsection{Evolution of other initial data: Qualitative features}
\label{s4.3.1}

We will highlight only those features which are qualitatively
different from the kinetic energy dominated bounce discussed in
detail in section \ref{s4.2}. Again, for concreteness, we will
focus on the case when $\phi_{\B}$ and $\dot\phi_{\B}$ are both
positive. Let us begin with the intermediate case $10^{-5}<f
\lesssim 1/\sqrt{2}$ where the kinetic energy still exceeds the
potential energy at the bounce but does not dominate it as in
section \ref{s4.2}. The super-inflation era is similar to that
described in section \ref{s4.2.1} but because $\dot\phi_{\B}$
is now lower, the phase lasts longer. The inflaton climbs up
the potential but change in its value is again small. The
Hubble parameter, on contrast, is again dynamical. However, the
post super-inflation dynamics exhibits significant differences.
For, now the value of $\phi_{\B}$ is higher and $\dot\phi_{\B}$
lower while, as before, $H$ assumes its largest value at the
end of super-inflation. Therefore, the coefficient of friction,
$\zeta=3H/2m$, is again large but there is less kinetic energy
to lose before reaching the turn-around., which is now reached
within $10-100 \sp$ after the bounce. Consequently, now the
change $\phi(t_{\rm TA})-\phi_{\B}$ is negligible, a key
feature not shared by regime (i).
%
%
%

Next, let us examine the $f \gtrsim 1/\sqrt{2}$ where the potential
energy is greater than the kinetic energy. Now, the LQC effects
dominate for an even longer time. Again, because $\dot\phi >0$, the
inflaton climbs up the potential but turns around earlier and
earlier and the super-inflation phase lasts longer and longer as $f$
increases. Now the turn around ($\dot\phi =0$) will occur
\emph{during super-inflation!} The change $(\phi (t_{\rm TA})
-\phi_{\B})$ is even more negligible because the kinetic energy at
the bounce is
lower than that in the $f<0.5$ case. 
The slow roll conditions are easily met soon after turn-around.
A difference from this phase of slow roll and that for
$f>1/\sqrt{2}$ is that $H$ continues to grow during the slow roll
because we are still in the super-inflation phase. There is an
enormous number of slow roll e-foldings already in the deep Planck
regime where the matter density is greater than half the critical
density.


\subsubsection{Compatibility with the WMAP data}
\label{s4.3.2}

In the extreme kinetic energy domination we could carry out
numerical simulations starting from the bounce until the end of the
desired slow roll. For $f>0.1$, this becomes quite difficult even
with the truncation error as small as one part in $10^{14}$ because
the super-inflation phase can become so long that the required
evolution spans $\sim 10^{14} \sp$ or more. Therefore, to maintain
the precision we had in the kinetic dominated case, another line of
reasoning is necessary.

\begin{figure}[ht]  \includegraphics [width=6in] {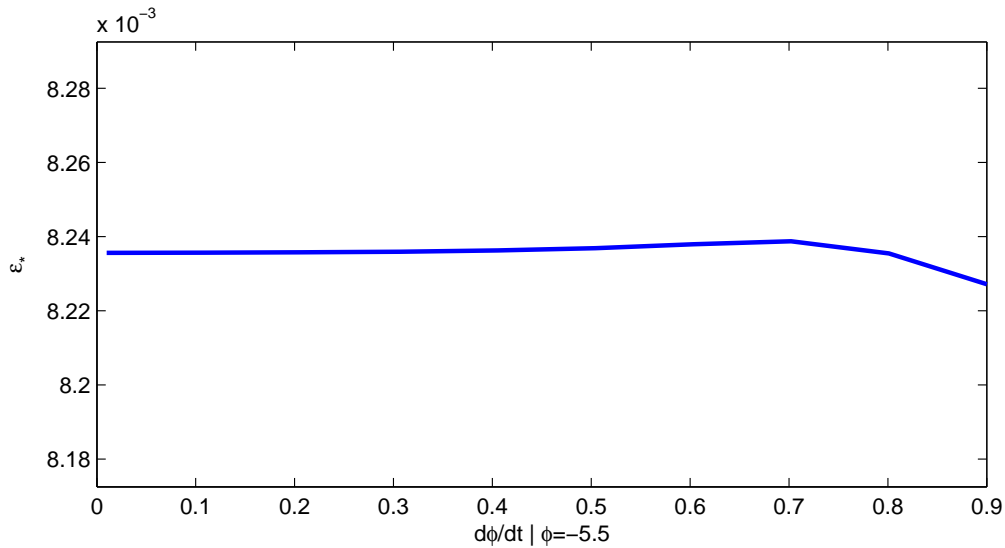}
\caption{ WMAP data implies that, for a quadratic potential,
at the time $t(k_{\star})$ at which the reference mode $k_\star$
exits the Hubble horizon, $|\phi| = 3.15\, mpl$ and $\epsilon =
8\times 10^{-3}$ within error bars of $\sim 4.5\%$.
We wish to analyze whether generic dynamical trajectories pass
through this small neighborhood in the phase space. For reasons discussed in
section \ref{s4.3.2}, we consider trajectories with initial data
with a fixed, low value of $\phi$ (which we take to be $\phi =
-5.5\,\mpl$), allowing \emph{all} permissible values of $\dot\phi$.
The plot shows that in all the resulting dynamical trajectories, at
the time when $\phi = -3.15\, \mpl$, we have $\epsilon \approx 8
\times 10^{-3}$ with fluctuations less than $2\times 10^{-5}$. This
implies that \emph{all} LQC trajectories with $|f| \ge 7.35 \times 10^{-6}$
pass through the desired, small region of the phase space.} \label{Fig1}\end{figure}

Motivated by our result (\ref{suff}), in this subsection we consider
initial data at the bounce with $|\phi_{\B}| \ge 5.5\,
\mpl$.%
\footnote{This space has an overlap with the $f<10^{-5}$ case
already considered in detail in section \ref{s4.2}. This ensures
that there are no `gaps', i.e., we will cover the full space of
initial data at the bounce.}
We will first argue that the resulting dynamical trajectory
\emph{must pass} through the point with $|\phi| = 5.5\,\mpl$ with
\emph{some} kinetic energy. Since this is a low value of $|\phi|$,
the inflation will reach the WMAP value $|\phi| = 3.15\,\mpl$ rather
quickly. So it is feasible to carry out a numerical simulation for
dynamics between $|\phi| = 5.5\,\mpl$ and $|\phi| = 3.15\, \mpl$
with the desired, high degree of accuracy. The question for numerics
is then: Is the value of $\epsilon$ at the time when $|\phi| =
3.15\, \mpl$ close to the WMAP value $\epsilon= 8\times 10^{-3}$
within the 4.5\% error bars? For initial data at the bounce for
which $|\phi_{\B}| \gg 5.5 \mpl$, the super-inflation phase will
last very long time and, had we begun the simulation at the bounce,
we would have lost accuracy by the time the inflaton reaches $|\phi|
= 3.15\, \mpl$. We avoid this problem by giving an \emph{analytic
argument} that the inflation \emph{must} reach $|\phi| = 5.5 \,
\mpl$ and carry out numerical calculations for dynamics only between
$|\phi|= 5.5\, \mpl$ and $|\phi| = 3.5\, \mpl$.

Let us begin with the negative branch of the $\phi_{\B}$ space,
i.e., first consider the case when $-5.5\, \mpl \ge \phi_{\B} \ge
-|\phi_{\rm max}| = -7.47\times 10^{5}\,\mpl$. In this case,
immediately after the bounce the inflaton will roll down the
potential. Intuitively, it may seem obvious that in this descent it
must encounter the point $\phi = -5.5\, \mpl$ with \emph{some}
kinetic energy. But a priori there are two possibilities that may
prevent this occurrence. First, we have to examine the possibility
that the inflaton may come to rest for some value $\phi=\phi_o <
-5.5\, \mpl$ and just stops there. But the equation (\ref{inflaton})
satisfied by the inflaton implies that at the instant $\dot\phi
=0$,\, $\ddot \phi = -m^2\phi_o >0$ (since $\phi_o \not=0$).
Therefore the inflaton cannot just stay at $\phi_o < 0$. A more
subtle possibility is that the inflaton asymptotically approaches
$\phi_o < -5.5\, \mpl$, with $\dot\phi$ approaching zero, but never
actually reaches it in a finite time. That is, the limiting value of
$\dot\phi$ could be zero at $\phi=\phi_o$ but this could happen at
time $t_o=\infty$. But by integrating (\ref{inflaton}) between the
bounce time $t_{\B}$ and the hypothetical time $t_o$ at which
$\dot\phi$ is to vanish, it is easy to show that:
\be  t_o - t_{\B} < \f{|\dot\phi_{\B}|}{m^2 |\phi_{\B}|}  \ee
so that $t_o$ is necessarily finite. Thus, if $\phi_{\B} < -5.5
\mpl$ (and by assumption $\dot\phi_{\B} >0$), the inflaton
\emph{must} slide down the potential and reach the value $\phi =
-5.5\, \mpl$ with \emph{some} kinetic energy.

Next, consider the case when $\phi_{\B} >0$. In this case, our focus
will be on initial data at the bounce with $\phi_{\B} \ge 5.5\,
\mpl$. Since we again have $\dot\phi_{\B}>0$, initially the inflaton
now \emph{climbs up} the potential. However, it again follows from
(\ref{inflaton}) that it reaches $\dot\phi=0$ in a finite time $t_o$
and at that time $\ddot\phi = -m^2\phi <0$ (since $\phi$ is now
positive). So this is the turn around point and the inflaton rolls
down the potential. Again, by the same reasoning as before, as it
rolls down, the inflaton cannot come to rest for a value $\phi \not
=0$. Therefore as it rolls down, it must pass through the point
$\phi= 5.5\, \mpl$, for \emph{some} value of $\dot\phi$. To
summarize, dynamical trajectories for \emph{every initial data at
the bounce with} $|\phi_{\B}| \ge 5.5\, \mpl$ \emph{eventually
encounter a point at which} $|\phi_{\B}| = 5.5\, \mpl$.

We used high precision numerics to analyze the LQC dynamics
following this event. The value of $\dot\phi$ at this event can be
arbitrary. We sampled the full range, $(\dot\phi=0,\, \dot\phi=
0.90\,\mpl^2)$, first using uniformly distributed \emph{thousand}
data points and then logarithmically distributed \emph{thousand}
points. Using $\phi= -5.5\,\mpl$ and each of these values of
$\dot\phi$ as initial data solved the full set of LQC equations
numerically. The key questions then are: i) Do all these dynamical
trajectories eventually pass through a phase space point at which
$\phi= -3.15 \, \mpl$; and, ii) Within the WMAP error bars, which of
them have $\epsilon = 8\times 10^{-3}$ at the time when $\phi$
assumes the value $3.15\,\mpl$? As one would expect from the
analytical considerations discussed above, the answer to the first
question is in the affirmative. As the inflaton rolls down the
potential from $\phi = -5.5\, \mpl$, it necessarily encounters the
value $\phi = -3.15\mpl$. The answer to the second question is
summarized in Fig. \ref{Fig1}: \emph{in each of these two sets of
1000 simulations}, at the time the inflaton assumes the value $\phi
= -3.15\,\mpl$, the slow roll parameter $\epsilon$ takes values in
the range
\be  \epsilon = 8\times 10^{-3}\, \pm\, 2 \times 10^{-5}\, . \ee
That is, each of these trajectories passes through the small portion
of the phase space compatible with the WMAP data.

Finally, let us consider the complementary case with initial data at
the bounce satisfying $\phi_{\B} > 5.5\, \mpl$. In this case, as
discussed above, the inflaton first rises up the potential and then
rolls down. Therefore now the numerical simulation should start with
$\phi = 5.5\,\mpl$ and $\dot\phi \in (-0.90\,\mpl, 0)$. But because of
the symmetry on the space of solutions noted in the beginning of
section \ref{s4.2}, these solutions can be obtained just by
reversing the sign of the solutions $\phi(t)$ obtained from
numerical evolutions starting from the initial data $\phi = -
5.5\,\mpl$ and $\dot\phi \in (0, \, 0.90\,\mpl)$. Thus, all the
dynamical trajectories are compatible with the WMAP data also in
this case.

To summarize, combining the results of sections \ref{s4.2.3} and
numerical simulations just discussed, we can conclude that the LQC
dynamical trajectories resulting from \emph{any} initial data at the
bounce surface with $|\phi_{\B}| \ge 5.5\, \mpl$ realizes the
constraints on values of fields at $t=t(k_{\star})$ imposed by the
WMAP data. Thus, it is only when the bounce is in the extreme
kinetic dominated regime that some of the dynamical trajectory
---namely those that violate (\ref{suff}) can fail to meet the
WMAP constraint.

\subsection{Probability of the desired slow roll in  LQC}
\label{s4.4}

We are now ready to combine results of sections \ref{s3.2},
\ref{s4.2} and \ref{s4.3} to calculate the a priori probability of
realizing the desired slow roll in LQC.

In the terminology of section \ref{s3.2}, the event $E$ of interest
is the passage of the dynamical trajectory through the small region
in the phase space singled out by the WMAP data and our task is to
find the \emph{relative} volume of the region $R(E)$ of the space of
solutions in which $E$ occurs. The calculation requires a normalized
measure. In section \ref{s3.2} we found that, thanks to a suitable
gauge fixing, each equivalence class of physically distinct LQC
solutions can be characterized completely by the value $\phi_{\B}$
of the inflaton at the bounce surface. Since $b\lambda =\pi/2$ at
the bounce, (\ref{vol}) implies that the total measure on the space
$\S_o$ of physically distinct solutions is given by
\be N = \int_{\S_o} \hat\omega = \int_{-\phi_{\rm max}}^{\phi_{\rm
max}}\, \big( \f{3\pi}{\lambda^2} - 4 \pi^2\gamma^2 m^2
\phi^2\big)^{\f{1}{2}}\, \dd\phi = \f{3\pi}{4\gamma\lambda^2m} \ee
where we have used the fact that we have a quadratic potential and
where, for definiteness we have set the volume $v_o$ at the bounce
to $1$. (Recall that $v_o$ simply rescales the measure by a constant
and therefore drops out in the calculation of probabilities.) The
probability $P(E)$ of the occurrence of our event $E$ is then given
by (\ref{prob}):
\be P(E) = \frac{1}{N}\, \int_{\I(E)} \big( \f{3\pi}{\lambda^2} - 4
\pi^2\gamma^2 m^2 \phi^2\big)^{\f{1}{2}}\, \dd\phi \ee
where $\I(E)$ is the interval on the $\phi_{\B}$-axis corresponding
to the physically distinct LQC solutions in which $E$ occurs. (For
details, see section \ref{s3.2}.)

In sections \ref{s4.2} and \ref{s4.3} we found that a
\emph{sufficient} condition for the event $E$ to occur is that the
initial data of the solution at the bounce should satisfy
\be \phi_{\B} \,\, \not\in\,\, [-5.5\,\mpl,\, \,.94\,\mpl]\quad
{\hbox{\rm or, equivalently,}}\quad f\,\, \not\in\,\, [-7.35\times
10^{-6},\,\, 1.25\times 10^{-6}] \ee
Therefore the probability that $E$ does \emph{not} occur is bounded
by the length of the interval $[-5.5\,\mpl,\, .94\,\mpl]$ in the
$\phi_{\B}$-axis w.r.t. the measure given by the 1-form
$\hat\omega$:
\ba P({\hbox{\rm E {\it not} realized}})\,&\le&\,\f{1}{N}\,
\big(\int_0^{\phi_+} +\int_0^{\phi_ -} \big)\, \big(
\f{3\pi}{\lambda^2} -
4 \pi^2\gamma^2 m^2 \phi^2\big)^{\f{1}{2}}\, \dd\phi \nonumber\\
&\le& \f{1}{\pi} (f_+\, \sqrt{1-f_+^2} + \sin^{-1} f_+)\, + \,
\f{1}{\pi} (f_-\, \sqrt{1-f_-^2} + \sin^{-1} f_- )\nonumber\\
&\lesssim&\,\,  2.74 \times 10^{-6}\ea
where $\phi_+ = .94\,\mpl,\, \phi_- = 5.5\, \mpl$ and $f_\pm =
\phi_\pm/\phi_{\rm max}$. Thus, the probability that the desired
slow roll does \emph{not} occur in an LQC solution is \emph{less
than three parts in a million}. Hence a great deal of fine tuning
would be necessary to avoid the slow roll inflation that meets the
WMAP constraints.

For simplicity, throughout our analysis, we set the cosmological
constant $\Lambda$ to be zero. But it is completely straightforward
to include it since it just shifts the zero of our potential. Since
the sign of the observed cosmological constant is positive, for any
given value of $\phi$, its inclusion would have the effect of
increasing the fraction of the total energy density in the
potential. This in turn would \emph{lower} the values of
$|\phi_{\B}|$ that are needed in (\ref{suff}) to ensure that the
trajectory meets the WMAP constraint. Thus the probability of
achieving the desired slow roll would further increase. However,
since the observed value of $\Lambda$ is so small, this effect will
not be noticeable even at the level of numerical accuracy used in
this paper. We also focused on the spatially flat FLRW models
because they are observationally favored. But it is not difficult to
include spatial curvature. Again, given observational constraints,
we expect that this inclusion will not significantly alter our main
conclusions \cite{ds}. In these respects, the results are robust.

What about the choice of the inflaton potential? All our detailed
considerations hold only for a quadratic potential. Recall however
that a generic potential $V(\phi)$ is well-approximated by a
quadratic one near its minimum. Suppose $V(\phi)$ has a single
minimum $\phi_o$ and that $|V(\phi)-V(\phi_o) - (1/2) m^2\phi^2| \ll
V(\phi)- V(\phi_o)$ for all $\phi \in [-5.5\, \mpl,\, 5.5\,\mpl]$
where $m \approx 1.21\times 10^{-6}\,\mpl$. Then our considerations
will apply: As the inflaton slides down from $|\phi| = 5.5\mpl$ it
will enter a phase of slow roll inflation compatible with the WMAP
constraints. As an illustration, let us suppose that the potential
is a combination of a quadratic and a quartic parts:
\be V(\phi) = \f{1}{2} m^2\phi^2 + \f{\lambda}{4} \phi^4  \ee
Then, our assumption is met if $\lambda \ll 9.68\times 10^{-14}$.
For a rough comparison, note that the WMAP bound for a pure quartic
potential is $4.4\times 10^{-14}$. Thus, it is not unreasonable to
expect that, at a qualitative level, our conclusions on
probabilities will continue to hold for a wide class of physically
interesting potentials.

\section{Discussion}
\label{s5}

In the main body of this paper we analyzed inflation in the
observationally most interesting case of the k=0 FLRW cosmologies
using the framework of LQC. Our goal was two-fold. First, assuming a
quadratic potential, we examined in detail the effective LQC
dynamics in presence of an inflaton with standard potentials.
Second, we used this dynamics to calculate the \emph{a priori}
probability of realizing the desired slow roll inflation that is
compatible with the WMAP data.

Let us begin with dynamics. Since the big bang singularity is
replaced by the big bounce, and since all physical fields are
regular at the bounce, we could specify initial conditions at the
bounce surface and explore the resulting dynamics. We allowed
\emph{all possible} initial data, subject only to the Hamiltonian
constraint. We found that the subsequent dynamics is sensitive to
the distribution of energy density at the bounce between kinetic and
potential parts. In all cases, there is a universal super-inflation
phase that follows the bounce; indeed it persists even when there is
no potential at all. For concreteness, let us focus on the case when
$\dot\phi$ and $\phi$ are both positive at the bounce so the
inflaton is climbing up the potential.

We first discussed in detail the case in which less than one part in
$10^{10}$ of the total energy density at the bounce is in the
potential because this is the only case in which some of the
solutions can fail to exhibit the desired slow roll. In this case,
the super-inflation phase is very short lived. However, it is also
extremely dynamic: during the fraction of a Planck second of
super-inflation, the Hubble parameter increases dramatically from
zero to its \emph{maximum allowed value}, $0.93 \,\mpl$. Because $H$
is so large at the end of super-inflation, the friction term in the
equation of motion of the inflaton is also large whence the inflaton
loses kinetic energy as it climbs up the potential. After $\sim 10^4
\,\sp$, the potential energy equals kinetic energy and then
dominates it (where, as before, $\sp$ denotes Planck seconds). After
another $\sim 10^4 - 10^5\,\sp$, \, $\dot\phi$ vanishes, the
inflaton turns around and starts climbing down the potential. In a
`majority' of solutions ---but not all--- it soon enters the desired
slow roll that is compatible with the 7-year WMAP observations
\cite{wmap}.

In the intermediate case, the kinetic energy is still greater than
the potential at the bounce but does not overwhelm it. Now dynamics
undergo the same qualitative phases. However, the super-inflation
phase lasts a bit longer and the turn around occurs much sooner,
only about $10-100 \sp$ after the bounce. This is because the
initial kinetic energy is smaller than that in the first case while
the strength of the friction term is the same because the Hubble
parameter again takes its maximum value at the end of
super-inflation. Because the turn around occurs much sooner, in
contrast to the situation in the extreme kinetic domination, the
increase in the value of $\phi$ between the bounce and the
turn-around is very small. Finally, if the potential energy density
exceeds the kinetic energy density at the bounce, dynamics is very
different. Now the LQC effects dominate in the sense that the
super-inflation phase lasts longer and its duration increases
rapidly as the fraction of total energy in the potential increases.
Furthermore, the turn around occurs already during super-inflation!
Our first goal was to present these qualitative differences in the
pre-slow-roll dynamics. 

The WMAP data puts very strong constraints on the values of fields
at the time $t(k_{\star})$ when the mode with wave number
$k_{\star}$ exits the Hubble radius during the desired phase of
inflation. The key question is: Are `most' dynamical trajectories
such that these constraints are met at \emph{some} time during the
post-bounce evolution? We showed that the answer is in the
affirmative unless the bounce is kinetic energy dominated as in
section \ref{s4.2}. Even in this case, the WMAP constraints are met
unless fraction of the total energy at the bounce which is in the
potential is less than $7.35 \times 10^{-6}$. In this precise sense
the tiny region of phase space selected by the WMAP data serves as
an \emph{attractor} to LQC dynamics. This may not seem surprising
because inflationary trajectories are known to be attractors also in
general relativity. However, in LQC there are strong quantum gravity
effects in the super-inflation phase as well as in the phase that
follows immediately after super-inflation. These could well have
spoiled the attractor behavior and a large fraction of the data
\emph{specified at the bounce} could well have given rise to
trajectories that steer away from the WMAP region. That this does
not happen is noteworthy.

The attractor behavior is a qualitative feature of dynamics. It does
not provide a sharp \emph{quantitative} estimate on the likelihood
of realizing the desired slow roll. To obtain this estimate, one
needs a \emph{measure} on the space of (physically distinct)
solutions, such that the total volume of this space is finite. As
explained in section \ref{s1}, a natural strategy \cite{ghs,dp,hp}
is to use the Liouville measure on the phase space. This leads to
\emph{a priori} probabilities of events, such as the occurrence of
the desired slow roll. However, volume of the 2-dimensional space
$\S$ of solutions with respect to the resulting $\dd\hat{\mu}_{\rm
L}$ is infinite. But this infinity is physically spurious: It occurs
because $\S$ is acted upon by a gauge group $\G$ and the length of
1-dimensional orbits of $\G$ is infinite. For a large class of
potentials the space $\S/\G$ of physically distinct solutions is in
fact a \emph{bounded} interval of the real line. Furthermore there
is a natural 1-form $\omega_\alpha$ on $\S$ (constructed from the
symplectic structure and the gauge vector field $G^a$) which is
everywhere orthogonal to $G^\alpha$. So it is a natural candidate to
induce a volume element on the 1-dimensional space $\S/\G$.
Unfortunately, it does not project down to $\S/\G$ (because
$\Lie_G\, \omega_\alpha = \omega_\alpha \not=0$). Therefore one is
led, instead, to `fix a gauge', i.e., lift $\S/\G$ to a suitable
1-dimensional cross-section in $\S$ and carry out integration there.
The problem in general relativity is that there is no natural family
of lifts and the volume element on $\S/\G$ depends heavily on one's
choice. Thus, there is an inherent ambiguity in the calculation of
probabilities. Thanks to the existence of the bounce surface, in LQC
one does have a natural family of lifts and the probability of
occurrence of physical events is independent of the
choice.%
\footnote{In this construction, the analog of the bounce surface in
general relativity would be the singularity itself. It is far from
obvious how to work at the singularity although ideas proposed in
\cite{foster} may provide a natural mathematical construction.
However, a priori, the physical meaning of such a construction would
be obscure because one would have to assumes that Einstein's
equations, without quantum corrections, are valid all the way to the
singularity.}
We used this choice to calculate the a priori probability of
realizing the desired slow roll, compatible with the WMAP data. We
found that the probability is \emph{greater than} 0.999997 in LQC.
As emphasized in section \ref{s1}, these are just `bare
probabilities' and better estimates of this occurrence will require
astutely chosen physical inputs. This will require not only a more
complete theory \cite{hw2}, but a deep understanding of that theory
to separate essential inputs from non-essential ones. However, since
the `bare probability' is so close to 1, it is a huge burden on any
theory to come up with new inputs that significantly change the
answer.

We emphasize, however, that our results should not be interpreted to
mean that sufficiently long slow roll inflation is inevitable in
LQC: it is inevitable \emph{only} under the additional assumption
that there is a phase in which matter density is dominated by that
of a scalar field in a suitable potential. So far LQC has not
provided a mechanism to create either the scalar field or the
potential. Although there have been intriguing suggestions that this
may naturally occur if one promotes the Barbero Immirzi parameter
$\gamma$ of loop quantum gravity to a dynamical field \cite{ty},
very substantial work is needed to have confidence that they can be
transformed into a concrete, internally complete and viable
scenario.

The primary purpose of our analysis was to understand what LQC has
to say about inflationary scenarios in the early universe. But it is
instructive to compare and contrast the predictions of LQC with
those of general relativity, discussed in the literature. This leads
us to several distinct points. First, there is some controversy as
to which `problems' inflation solves and which it does not
\cite{hw1,klm,hw2}. Our analysis does not shed any new light on
these issues. Rather, as mentioned above, our focus is on the
likelihood of the desired slow roll inflation. Second, while some of
the literature \cite{ghs,dp,hp,gt} uses the Liouville measure on the
space of solutions to compute the a priori probability, in
\cite{klm} (and in the earlier literature cited therein) a different
measure is used. The likelihood, of course, can and does depend
sensitively on the choice of the measure. In addition, since the
choice made in \cite{klm} is not preserved by the Hamiltonian flow,
a choice of time slice is made. Therefore, there is some discussion
in the literature on the dependence of the final results on the
choice of the measure and the time slice used in the evaluation of
probabilities (see, e.g., \cite{hw1,hw2,gt}). The Liouville measure
is `canonical' and, furthermore, preserved under time evolution.
However, as explained above, the total measure on the space of
solutions $\S$ is infinite and, to extract meaningful probabilities,
one must introduce an additional structure, a suitable lift of
$\S/\G$ into $\S$. Conceptually, this is equivalent to fixing a
`time gauge,' e.g. by working with the initial data of the solution
at the instant at which the matter density is a given constant,
$\rho_o$. In this sense, there is a common limitation. In \cite{klm}
(and in earlier works dating back to \cite{bkgz}) one uses the time
slice at Planck time (but still works with equations of general
relativity), while in \cite{gt} it is argued that one should use a
slice corresponding to a \emph{much} later time. Even if one were to
decide to use the Liouville measure, a priori probabilities do
depend sensitively on this choice because of the nature of
inflationary dynamics \cite{ck}. From our perspective, since there
is no `canonical' choice of time slice in general
relativity, calculations of probabilities have an inherent ambiguity.%
\footnote{There is another difference between references \cite{klm}
and \cite{gt}: while \cite{klm} focuses on the (phenomenologically
favored) k=0 model, \cite{gt} focuses on the (spatially closed) k=1
model. However, since the strategy used in \cite{gt} to `regularize'
the Liuoville measure is strongly motivated by the gauge
considerations in the k=0 case, this difference is less significant
for the drastic disparity in the final results.}
In LQC, by contrast, since the bounce surface provides a canonical
`time slice' the ambiguity can be naturally resolved. Finally,
recently there have been several interesting discussions of
inflationary scenarios in LQC (see, e.g.,
\cite{gns,gb,gcbc,gbcm,mcgb,barrau}). However, the primary focus of
all but one of these papers is on phenomenological and observational
issues rather than on measures and  calculations of probabilities.
The one exception is \cite{gns}. However, in that reference,
super-inflation (as well as the bounce) was ignored.

We will conclude with an observation on dynamics immediately
following the bounce. In the standard inflationary scenarios, matter
density is some 11-12 orders of magnitude smaller than $\rho_{\rm
Pl}$ at the onset of inflation and from the quantum equations of
full LQC we know that the use of quantum field theory in curved
space-times is fully justified in this regime. However, there are
several conceptually important issues which require an understanding
of dynamics \emph{before} this era is reached. Perhaps the most
important among them is the issue of the initial state of quantum
fields representing perturbations. As explained in section \ref{s1},
currently the state is simply postulated to be the Bunch-Davis
vacuum for the relevant modes (which have co-moving wave numbers in
the range $(k_o, \,\, \sim 200 k_o)$). While this assumption is
physically motivated, it is nonetheless important to arrive at this
state starting from more fundamental initial conditions. In general
relativity, these conditions would have to be specified on the
singularity and furthermore we know that we cannot reliably use
equations of general relativity in the Planck regime. LQC on the
other hand is based on loop quantum gravity, a candidate theory of
full quantum gravity. Therefore, not only is there a basis to trust
its quantum equations but, as we saw in section \ref{s2}, they have
already provided us a wealth of new information on physics at the
Planck scale. Therefore, there is good motivation to use quantum
field theory on the cosmological \emph{quantum space-times} of LQC
\cite{akl} to study the initial conditions and evolution of quantum
fields representing perturbations from the big bounce until the
onset of inflation. Is there perhaps a natural initial condition we
can impose on the quantum state of these perturbations at the bounce
using e.g., ideas developed in \cite{cmov} and/or exploiting the
fact that the Hubble parameter vanishes there? What would such an
initial state evolve to at the onset of slow roll? Would it be
sufficiently close to the Bunch-Davis vacuum (for the modes of
interest) to be phenomenologically viable, or, would it be so
different that it is already ruled out observationally? If it close,
is it perhaps too close to be observationally indistinguishable from
the Bunch-Davis vacuum or is it sufficiently different to lead to an
observational test of LQC? These questions are being currently
analyzed using the detailed dynamics of the LQC inflationary
space-times presented in section \ref{s4.2} \cite{aan}.

Finally, because almost all LQC solutions are compatible with the
WMAP data, one might first think that the chances of constraining
quantum gravity from observations of the early universe are very
small. However this is not correct: to make contact with these
observations one needs not only the `background' space-time but also
quantum fields representing perturbations off this background.
Considerations of the previous paragraph suggest a concrete
direction to confront quantum field theory on the LQC quantum
space-times with observations. Such a confrontation could well lead
to interesting, testable predictions of loop quantum gravity and/or
constraints on this theory.

 \section*{Acknowledgments}
We would like to thank Ivan Agullo, Alejandro Corichi, Neil Turok
and William Nelson for discussions. This work was supported in part
by the NSF grant PHY0854743 and the Eberly and Frymoyer research
funds of Penn State.

\end{document}